\documentclass[trackchanges]{aastex7}
\DeclareUnicodeCharacter{02BC}{\textquotesingle}

\begin{document}

\title{Identifying Dust-lane Spheroidal Galaxies in DESI Legacy Imaging Surveys Using Semi-Supervised Methods}

\author[orcid=0009-0009-1617-8747,sname=Luo]{Zhijian Luo}
\affiliation{Shanghai Key Lab for Astrophysics, Shanghai Normal University, Shanghai 200234, People’s Republic of China}
\email[show]{zjluo@shnu.edu.cn} 


\author[gname=Jianzhen, sname=Chen]{Jianzhen Chen} 
\affiliation{Shanghai Key Lab for Astrophysics, Shanghai Normal University, Shanghai 200234, People’s Republic of China}
\email[show]{jzchen@shnu.edu.cn}

\author[0009-0000-0690-5562,sname=Pei,gname=Wenxiang]{Wenxiang Pei}
\affiliation{Shanghai Key Lab for Astrophysics, Shanghai Normal University, Shanghai 200234, People’s Republic of China}
\email{wxpei@nao.cas.cn}

\author[0000-0001-8244-1229,sname=Xiao,gname=Hubing]{Hubing Xiao}
\affiliation{Shanghai Key Lab for Astrophysics, Shanghai Normal University, Shanghai 200234, People’s Republic of China}
\email{hubing.xiao@outlook.com}

\author[0000-0002-2326-0476,sname=Zhang,gname=Shaohua]{Shaohua Zhang}
\affiliation{Shanghai Key Lab for Astrophysics, Shanghai Normal University, Shanghai 200234, People’s Republic of China}
\email{zhangshaohua@shnu.edu.cn}

\author[0009-0001-5320-1450,gname=Qifan]{Qifan Cui}
\affiliation{Shanghai Key Lab for Astrophysics, Shanghai Normal University, Shanghai 200234, People’s Republic of China}
\email{qfcui@nao.cas.cn}

\author[gname=Chenggang]{Chenggang Shu}
\affiliation{Shanghai Key Lab for Astrophysics, Shanghai Normal University, Shanghai 200234, People’s Republic of China}
\email{cgshu@shao.ac.cn}

\begin{abstract}

Dust-lane spheroidal galaxies (DLSGs) are unique astrophysical systems that exhibit the morphology of early-type galaxies (ETGs) but are distinguished by prominent dust lanes. Recent studies propose that they form through minor mergers between ETGs and gas-rich dwarf galaxies, offering a window into the interstellar medium (ISM) of ETGs and star formation triggered by small-scale interactions. However, their rarity poses a challenge for assembling large, statistically robust samples via manual selection. To overcome this limitation, we employ GC-SWGAN, a semi-supervised learning method developed by \citet{2025ApJS..279...17L}, to systematically identify DLSGs within the DESI Legacy Imaging Surveys (DESI-LS). The methodology involves training a generative adversarial network (GAN) on unlabeled galaxy images to extract morphological features, followed by fine-tuning the model using a small dataset of labeled DLSGs. In our experiments, despite DLSGs constituting only $\sim$ 3.7\% of the test set, GC-SWGAN achieves remarkable performance, with an 87\% recall rate, 84\% accuracy, and an F1 score of 86\%, underscoring its efficacy for DLSG detection. Applying this model to $\sim$ 310,000 DESI-LS galaxies that meet the criteria $m_r < 17.0$ and $0.01 < z < 0.07$ we compile the largest catalog of DLSG candidates to date, identifying 9,482 dust-lane ETGs. A preliminary analysis reveals that these DLSGs exhibit significantly redder $g-r$ colors and higher specific star formation rates compared to non-DLSGs. This catalog enables future studies of ISM properties in ETGs and the role of minor mergers in driving star formation in the nearby universe.

\end{abstract}

\keywords{\uat{Galaxies}{573} --- \uat{Astronomy data analysis}{1858} --- \uat{Astronomy image processing}{2306} --- \uat{Convolutional neural networks}{1938} --- \uat{Ground-based astronomy}{686} --- \uat{Computational astronomy}{293}}


\section{Introduction} \label{sec:intro}


Early-type  galaxies (ETGs) are one of the primary types of massive galaxies in the universe. Their stellar populations not only record the process of mass accumulation throughout cosmic history, but also preserve detailed fossil records of their progenitor galaxies. Although extensive studies have shown that most of these galaxies’ stellar populations ($>$ 80\%) are ancient and rapidly formed during high-redshift periods (z $>$ 1) \citep{1992MNRAS.254..589B,1996MNRAS.280..167J,1998ApJ...508L.143B,1999MNRAS.302..537T,2000AJ....119.1645T,2000AJ....120..165T,2003AJ....125.1882B,2004ApJ...608..752B,2007ApJ...665..265F}, recent studies using ultraviolet (UV) photometry have revealed that they exhibit widespread, low-level, ongoing star formation at recent epochs (z $<$ 1) \citep{2004ApJ...601L.127F,2005ApJ...619L.111Y,2007MNRAS.376.1021J,2007ApJS..173..619K,2008MNRAS.388...67K,2010ApJ...714L.290S,2011ApJ...727..115C,2018ApJ...864..123M,2024MNRAS.531.2223P}. This star formation is primarily the result of minor mergers with gas-rich dwarf galaxies \citep{2009MNRAS.394.1713K,2011MNRAS.411.2148K,2012ApJ...746..162N,2012ApJ...749...53R}.

With mounting evidence indicating the ubiquity of star formation driven by minor mergers, the study of interstellar matter in ETGs is becoming an increasingly popular focus in the field of astronomy. Although ETGs have traditionally been regarded as dry and passively evolving systems, numerous studies have shown that most ETGs in the nearby universe contain dust \citep{1980ApJ...241..969T,1981MNRAS.196..747H,1985MNRAS.214..177S,1988AJ.....95..422E,1989ApJS...70..329K,1995A&A...298..784G,1995AJ....110.2027V,1996ApJ...460..271K,1996MNRAS.280..167J,1997AJ....114.1771F,2000AJ....120..123T,2001AJ....121.2928T,2007MNRAS.377.1795C,2008A&A...479..669C,2011MNRAS.414..940Y,2012ApJ...748..123S,2013MNRAS.431.1929A}. Moreover, the dust mass in these galaxies cannot be entirely attributed to the loss of stellar mass within the galaxies, suggesting that a significant portion of the interstellar matter in ETGs may originate from external sources \citep{1993ApJ...419..544S,1998A&A...338..807M,2010A&A...519A..40A,2011MNRAS.417..882D,2012ApJ...748..123S,2019MNRAS.489L.108D,2022MNRAS.517.5524G}. These findings provide a new perspective for understanding the formation and evolution of ETGs and offer important clues for in-depth studies of the physical processes within these galaxies.

Interstellar matter is a crucial raw material for star formation, and its properties and distribution play a vital role in the evolution of galaxies. However, our understanding of interstellar matter, especially dust in ETGs, remains limited. Dust-lane spheroidal galaxies (DLSGs) are a rare class of celestial objects that are believed to be remnants of recent minor mergers \citep{1981MNRAS.196..747H,2002AJ....123..729O,2008MNRAS.386L..82M,2010A&A...514A..57S,2012MNRAS.423...59S,2015MNRAS.449.3503D,2022MNRAS.517.5524G,2024ApJS..274....3G}. These systems are ideal for investigating the interstellar medium (ISM) in ETGs and for studying the minor merger processes that drive late-stage star formation.

Although detailed studies of DLSGs are of great significance, early research has been constrained by a lack of statistical samples due to their rarity. In recent years, \citet{2012MNRAS.423...49K} (hereafter K12) conducted an extensive screening of DLSGs based on the Galaxy Zoo 2 (GZ2) project \footnote{\url{https://data.galaxyzoo.org/}}. From the GZ2 classifications of nearly 300,000 galaxies in the Sloan Digital Sky Survey (SDSS) DR7, about 19,000 galaxies were initially selected that were marked as having dust features by at least one GZ2 user. These galaxies were subsequently re-examined by S. Kaviraj and Y. S. Ting to confirm their dust characteristics, the morphology of the host galaxies, and the presence of disturbances. To ensure accurate detection of dust features, their study was restricted to a redshift range of $0.01 < z < 0.07$. Ultimately, a sample of 352 reliably classified DLSGs was obtained. This sample provides an important data foundation for subsequent studies of the local environment, dust characteristics, star formation, and active galactic nucleus (AGN) activity of DLSGs \citep{2012MNRAS.423...49K,2012MNRAS.423...59S}.

While previous studies have made significant progress in investigating the ISM in DLSGs, the statistical reliability of these findings remains constrained by limited sample sizes. Currently, the largest available DLSG sample, from K12, contains only a few hundred cases. This scarcity not only hinders our comprehensive understanding of the general properties of the ISM in DLSGs but also poses challenges in studying the formation and evolution of dust lanes, as well as their relationship with star formation and other galactic properties. Consequently, expanding the DLSG sample size substantially is crucial. An enlarged sample would enhance the statistical robustness of research outcomes, enabling more accurate insights into the universal characteristics of the ISM in ETGs.

However, identifying DLSGs in large-scale astronomical surveys presents significant challenges. These difficulties stem not only from the complex morphology of their dust lanes and the inherent difficulty in detecting such features but also from the extreme rarity of DLSGs within galaxy populations. As noted by K12, their identified sample of DLSGs represents approximately 4\% of early-type galaxies within the same redshift and magnitude range ($0.01 < z < 0.07$, $m_r < 17.0$), corresponding to only about 1\% to 2\% of the entire galaxy population. Traditional visual detection methods for identifying DLSGs among large numbers of galaxies are not only time-consuming and labor-intensive but also susceptible to subjective bias. Consequently, developing and implementing more efficient, accurate, and automated detection methods is critical for locating these rare objects.

In recent years, machine learning techniques have demonstrated significant potential in astronomical data analysis. Supervised learning methods, in particular, have been widely adopted for celestial classification tasks, including galaxy morphology classification (e.g., \citep{storrie1992morphological, banerji2010galaxy, dieleman2015rotation, walmsley2020galaxy}) and stellar/spectral classification (e.g., \citep{2015MNRAS.453..122S, 2020MNRAS.491.2280S, 2023NewA...9901965W}). However, these methods heavily depend on large volumes of high-quality labeled data, which limits their effectiveness for rare targets (such as DLSGs) where labeled samples are scarce.

Unsupervised learning, while advantageous for discovering hidden patterns in unlabeled data, faces its own challenges: rare objects like DLSGs are often underrepresented in training sets, making it difficult for models to learn their distinguishing features. As a result, unsupervised approaches frequently exhibit poor generalization and fail to achieve the classification accuracy needed for practical applications in rare-object identification.

Recently, \citet{2025ApJS..279...17L} proposed an innovative semi-supervised learning method called GC-SWGAN. This method combines the strengths of supervised and unsupervised learning, leveraging limited labeled data and abundant unlabeled data during co-training to significantly enhance the model's classification performance while substantially reducing dependence on labeled samples. Systematic experimental evaluations on the Galaxy10 DECaLS dataset \citep{2025ApJS..279...17L} demonstrate that GC-SWGAN can achieve effective and accurate classification even when labeled data are scarce. Compared to traditional fully supervised learning methods, this model requires only about 10\% to 20\% of the labeled data to reach comparable classification accuracy.

The significant advantages of GC-SWGAN make it particularly suitable for the identification of rare astronomical targets, paving a new technological path for the detection of such objects. Notably, \citet{chen2025detecting} successfully applied GC-SWGAN to DESI-LS survey data, identifying over 60,000 candidate ring galaxies using only a small number of labeled samples. This achievement not only validates the exceptional practical performance of GC-SWGAN but also demonstrates its efficiency and practicality when handling large-scale astronomical datasets.

In this study, we will systematically identify DLSGs in the DESI-LS using the GC-SWGAN model. The DESI-LS dataset, with its wide-area coverage, deep multi-band imaging, and high photometric quality, provides an ideal foundation for constructing a robust catalog of DLSG candidates. The adopted GC-SWGAN framework implements a two-stage optimization strategy to address the challenges of detecting rare astrophysical phenomena. In the initial unsupervised training phase, the model leverages a large sample of unlabeled galaxy images from DESI-LS to learn generalizable morphological features. This stage enables the algorithm to develop an understanding of galaxy structures without requiring labeled data. Building upon this foundation, the subsequent supervised fine-tuning phase employs advanced data augmentation techniques. This allows the model to achieve high classification accuracy even when working with limited labeled samples, effectively overcoming one of the primary challenges in rare object detection. 

Through this work, we establish an efficient, data-driven pipeline that represents a significant advancement over traditional manual classification methods. The framework provides a scalable solution for detecting rare astrophysical objects in large-scale surveys, while maintaining high accuracy and reproducibility. This approach not only enhances our ability to identify DLSGs but also lays the groundwork for future applications in other areas of astronomy where the detection of rare phenomena is critical.


The structure of this paper is organized as follows: Section \ref{sec:dataset} describes the datasets used in this study and their preprocessing procedures, including image cropping, normalization, augmentation, and the division of training and test sets. Section \ref{sec:method} presents the architecture of the semi-supervised GC-SWGAN neural network model used and its training process. Section \ref{sec:performance} showcases the training results of the GC-SWGAN model and its performance on the testing set. Section \ref{subsec:catalog} provides the catalogue of DLSGs generated by the model in the DESI-LS. Section \ref{sec:discuss} offers a detailed discussion and analysis of the generated DLSG catalogue, and explores the physical properties of the identified DLSGs in terms of their colors and star formation rates, utilizing data resources from the value-added public catalogue. Finally, Section \ref{sec:summary} summarizes the research findings and discusses their significance to the relevant fields, and directions for future research.

\section{Data Preparation} \label{sec:dataset}

The primary objective of this study is to train a semi-supervised model that accurately identifies DLSGs in publicly available DESI-LS images, thereby compiling a large-scale DLSG candidate catalog. This section details the selection and pre-processing of the model training samples from the DESI-LS and GZ2 datasets.

\subsection{Training Sample Selection} \label{subsec:sample_select}

To train a semi-supervised learning model for DLSG identification, two types of training datasets need to be prepared. The first is a labeled galaxy dataset, which includes both DLSG and non-DLSG samples. The quality and quantity of these labeled samples are crucial for model training. Typically, experienced astronomers or dedicated annotation teams are required to perform precise labeling of the samples to ensure that the model can accurately learn the distinguishing features between DLSGs and non-DLSGs. The second is an unlabeled galaxy dataset, which can be sourced extensively. Any number of galaxy images from the target domain can be selected for training.

For the DLSGs in the labeled training dataset, we chose the K12 galaxies as the initial sample. This sample originates from SDSS DR7 and was further refined based on visual classification results from the GZ2 project and the K12 authors. It is currently the largest known DLSG sample set.

GZ2 is a large-scale citizen science project that enlisted hundreds of thousands of volunteers to classify the morphologies of approximately 300,000 primary galaxies ($m_r < 17.0$) from SDSS DR7 \citep{willett2013galaxy}. The project has demonstrated exceptional performance in identifying peculiar objects, such as merging galaxies \citep{2010MNRAS.401.1043D,2010MNRAS.401.1552D}, ring galaxies \citep{2022MNRAS.513.1581W}, and dust lane galaxies (e.g., the K12 sample). These types of objects typically require visual inspection for reliable identification, and the massive classification dataset generated by GZ2 provides an ideal foundational sample for many studies.

The construction process of the K12 sample is briefly described as follows: The authors, S. Kaviraj and Y. S. Ting, initially selected approximately 19,000 galaxies from the GZ2 data that had been flagged by at least one user as potentially containing a dust lane (in GZ2, each galaxy received an average of over 50 independent classifications). Subsequently, they manually reviewed these galaxies to confirm the presence of dust lane features and further refined the selection to include only systems with early-type morphology.

Ultimately, they obtained a high-confidence sample of 352 DLSGs within the redshift range of $0.01 < z < 0.07$ (beyond $z > 0.07$, reliable detection of dust features becomes challenging). To assess the completeness of the sample, S. Kaviraj and Y. S. Ting examined 1,000 randomly selected remaining GZ2 galaxies to determine how many DLSGs might have been missed, i.e., not identified by at least one GZ user as containing a dust lane. Their test yielded no additional candidates, leading them to conclude that all DLSGs had been captured, suggesting the K12 sample is likely complete for systems with prominent dust lanes in GZ2 \citep{2013MNRAS.435.1463K}. Following further verification and supplementation, the final number of DLSGs published on the GZ2 website was 362.

\begin{figure} 
        \includegraphics[width=\textwidth]{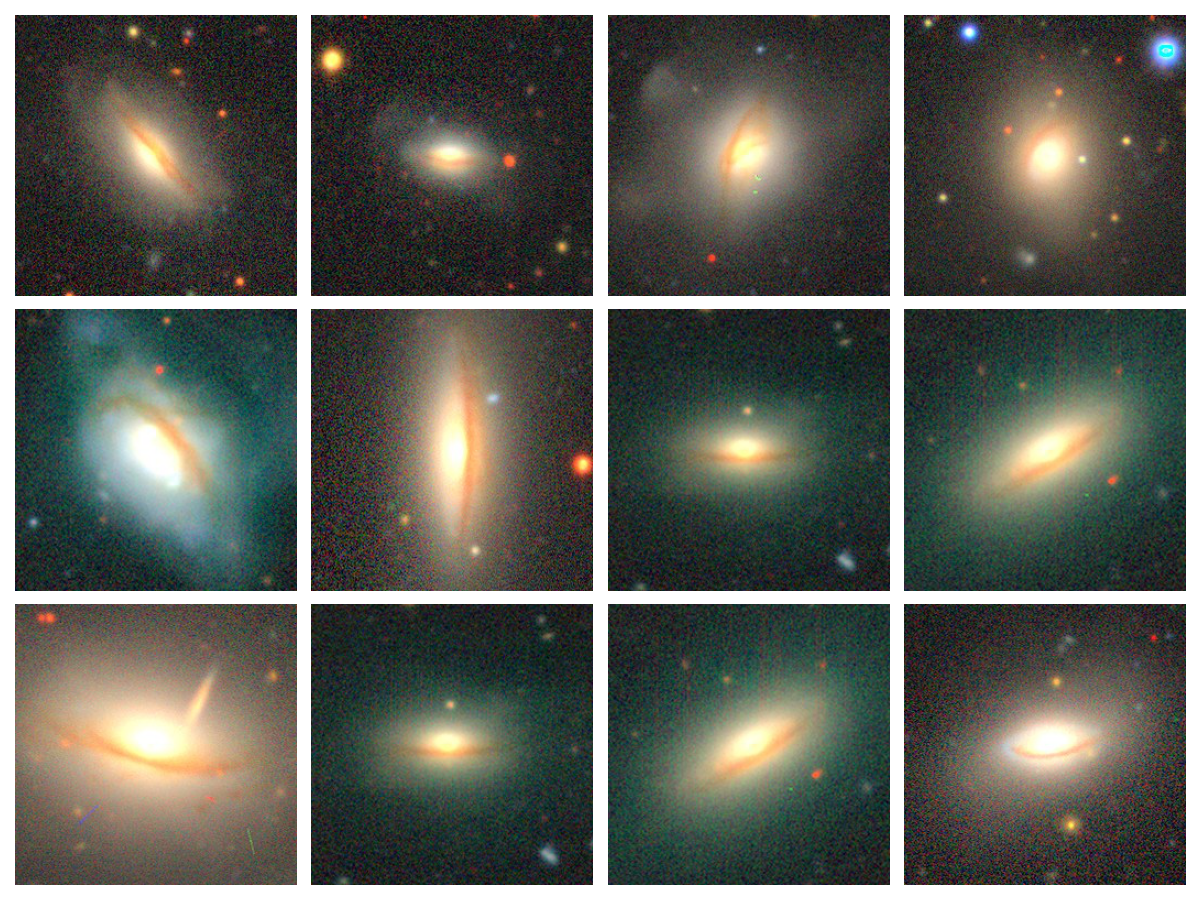} 
        \caption{Example DESI-LS color-composite images of 12 representative DLSGs from the training set. Each image is 256×256 pixels, corresponding to an angular size of approximately 67×67 arcseconds (scale: 0.262 arcsec/pixel). Images combine $g$-, $r$-, and $z$-band observations — assigned to blue, green, and red channels, respectively — from the DESI Legacy Imaging Surveys (BASS + DECaLS + MzLS).} 
        \label{fig:DLSGs_train} 
\end{figure}

Since dust lanes have a more pronounced effect on galaxy colors than on brightness, multi-band color images are more advantageous for detecting dust lane features in the model.
To achieve the objectives of this study, we downloaded the $g$, $r$, and $z$-band images of these 362 galaxies from the DESI-LS website and conducted multiple rounds of meticulous visual inspection. During the screening process, we excluded six galaxies with extremely faint dust lane features, resulting in an initial sample of 356 DLSGs for model training. Figure \ref{fig:DLSGs_train} displays some typical examples of these 356 DLSGs. As can be seen from the figure, these DLSGs exhibit distinct dust lane features.

The DESI-LS provide the essential imaging data for the Dark Energy Spectroscopic Instrument (DESI; \citealt{2019AJ....157..168D}) and support its cosmological research. DESI is mounted on the 4 m Mayall Telescope at Kitt Peak National Observatory, while DESI-LS comprises three complementary surveys: DECaLS, BASS, and MzLS. BASS (Beijing–Arizona Sky Survey) and MzLS (Mayall z-band Legacy Survey) jointly cover the northern sky. BASS uses the 2.3 m Bok Telescope at Kitt Peak for $g$ and $r$-band imaging \citep{2004SPIE.5492..787W}, whereas MzLS employs the 4 m Mayall Telescope for $z$-band observations \citep{2016SPIE.9908E..2CD}; together they are referred to as BASS/MzLS. DECaLS (Dark Energy Camera Legacy Survey) uses the 4 m Blanco Telescope at Cerro Tololo Inter-American Observatory to image the southern sky in $g$, $r$, and $z$ bands \citep{2015AJ....150..150F}. Together, BASS/MzLS and DECaLS provide approximately 14,000 square degrees of three-band ($g$, $r$, $z$) coverage. Additionally, the Dark Energy Survey (DES; \citealt{2016MNRAS.460.1270D}) also utilizes the Blanco Telescope’s DECam camera, and its 5,000 square degrees $g$, $r$, $z$ coverage of the South Galactic Cap is incorporated into the DESI-LS data release. In this study, we utilize the ninth data release (DR9) of the DESI Legacy Image Survey.

To construct a high-quality labeled training dataset, it is equally crucial to select non-DLSG samples properly in addition to correctly choosing DLSG samples. When filtering non-DLSG samples, we ensured that the same constraints were applied as those for DLSG samples, specifically, the apparent magnitude ($m_r < 17.0$) and the redshift range ($0.01 < z < 0.07$). This ensures the comparability between the two types of samples and the overall consistency of the dataset.

From the Galaxy Zoo 2 (GZ2) dataset, we randomly selected over 12,000 galaxies as initial non-DLSG candidates from a pool of approximately 100,000 galaxies that met the aforementioned criteria ($m_r < 17.0$, $0.01 < z < 0.07$) and did not include the known 356 DLSGs. For all these galaxies, we downloaded their $g$, $r$, and $z$-band images from the DESI-LS website.

To ensure the quality of these non-DLSG samples and maintain the purity of the dataset, two researchers (JZ and ZJ) conducted a thorough manual review of all these more than 12,000 galaxies. This review utilized DESI-LS color-composite images generated from the $g$, $r$, and $z$ bands and targeted five diagnostic features for DLSG identification: a) dominant early-type morphology; b) presence of ring-like or irregular dust lanes surrounding the galactic nucleus; c) well-defined lane boundaries; d) relatively uniform, faint, and reddish dust lanes without bright star-forming regions; e) consideration of line-of-sight inclination to exclude similar structures caused by external obscuration or instrumental artifacts. The examination revealed that fewer than 1\% (a total of 103 galaxies) of the sample — which should have been classified as DLSGs — were incorrectly labeled as non-DLSGs. This was largely due to the superior resolution and depth of DESI-LS compared to SDSS. We subsequently removed these 103 galaxies from the non-DLSG category and incorporated them into the K12 sample as confirmed DLSG training examples. This rectification process enhanced both the accuracy of our negative samples and the comprehensiveness of our positive DLSG training set.

Through this detailed data selection process, we have successfully established a high-quality labeled training dataset. This dataset comprises 459 DLSGs samples and 12,000 non-DLSGs samples, with DLSGs accounting for approximately 3.7\% of the total sample size. This proportion appropriately reflects the rarity of DLSGs in the broader galaxy population and provides a solid foundation for reliable model deployment in real astronomical surveys. 

For the unlabeled training dataset, the available resources are very abundant. Based on the constraints of apparent magnitude ($m_r < 17.0$) and redshift range ($0.01 < z < 0.07$), we randomly selected 20,000 galaxy samples from the Galaxy Zoo DESI database \citep{walmsley2023galaxy}. Subsequently, we downloaded their color images (covering the $g$, $r$, and $z$ bands) from the DESI-LS website as the unlabeled samples required for model training. The selected galaxy images exhibit a diverse range of morphological features, ensuring that the model can obtain rich morphological information. In addition, the DESI-LS website provides millions of unlabeled DESI galaxy images, offering great potential for expanding the unlabeled dataset. Such expansion will enable the model to utilize a broader range of training resources, further enhancing its generalization ability and recognition accuracy, thereby optimizing its overall performance.

\subsection{Data Preprocessing} \label{subsect:data_prep}

We applied a standardized preprocessing pipeline to all model training data. The preprocessing steps primarily included image cropping, pixel value normalization, dataset splitting, and training data augmentation.

First, regarding image cropping, all original galaxy images downloaded from the DESI-LS website had dimensions of 256×256 pixels, with each pixel corresponding to a resolution of 0.262 arcseconds. The primary galaxies were consistently centered in these images. To balance computational efficiency with model performance, we cropped these images from 256×256 pixels to 192×192 pixels. This strategy not only significantly reduced computational resource consumption (by approximately 32\%) but also preserved the main structural features of the galaxies. Studies have demonstrated that this reasonable size adjustment adequately maintains the principal structural characteristics of the vast majority of galaxies in our sample, thereby supporting the model's effective identification of key features in celestial targets \citep{radford2015unsupervised,2025ApJS..279...17L,luo2025cross}.

Secondly, regarding pixel normalization, the original images had pixel values ranging from [0, 255]. To normalize these values to the range of [-1, 1], we employed the following formula:

\begin{equation}
     x^* = \frac{x - 127.5}{127.5},
     \label{eq:normal} 
\end{equation}
where $x$ represents the original value of each pixel, and 127.5 is the median of the original range. This normalization approach not only helps stabilize gradient flow during model training but also preserves image detail features, thereby accelerating model convergence and improving training stability. This technique has been widely adopted in various deep learning tasks, including GANs and convolutional neural networks (CNNs) \citep{patro2015normalization,radford2015unsupervised,goodfellow2016deep,he2016deep}.

Next, regarding dataset partitioning, we adopted an 8:2 train-test split ratio for the labeled data. Specifically, the training set consists of 367 DLSGs and 9,600 non-DLSG samples, while the test set contains 92 DLSGs and 2,400 non-DLSG samples. For the unlabeled data, we fully utilized the 20,000 galaxy images described in Section \ref{subsec:sample_select}. Benefiting from the massive data continuously generated by the DESI survey, this dataset maintains significant potential for future expansion, which will provide a data foundation for further improving model performance.

Finally, regarding data augmentation, we exclusively applied augmentation techniques to the DLSG samples in the labeled training set. Given the severe class imbalance in the labeled training set (with DLSGs representing only ~3.7\% of samples), we implemented a dual augmentation strategy combining geometric transformations and SMOTE (Synthetic Minority Oversampling Technique; \citealt{chawla2002smote}) to enhance the model's recognition capability for minority classes and improve overall performance. Conversely, we intentionally avoided additional augmentation for both the non-DLSG samples in the labeled dataset and all samples in the unlabeled dataset. This deliberate choice was made because: (1) these samples already comprise a substantial number of galaxy images, and (2) they inherently possess sufficient morphological diversity to provide adequate training variability.

For geometric transformations, we utilize basic image operations to generate more diversified DLSG samples while ensuring the preservation of their critical morphological features. The three simple yet effective geometric transformations include horizontal flipping (randomly applying a horizontal mirroring effect), vertical flipping (randomly applying a vertical mirroring effect), and 90-degree rotation (rotating images either clockwise or counterclockwise by 90 degrees). These methods not only enhance the model’s ability to capture galaxy morphological symmetries but also do so without significantly increasing computational overhead, thereby efficiently improving overall performance.

To address the severe class imbalance in our training set, we additionally employed the SMOTE technique to augment the DLSG-class training samples. This widely-adopted oversampling method is particularly suitable for expanding limited minority class samples while maintaining their intrinsic characteristics. The algorithm operates by first randomly selecting a DLSG sample (denoted as $a$) and identifying its $k$-nearest neighbors within the same class. For each selected neighbor (sample $b$), the method generates new synthetic instances through linear interpolation along the line segment connecting $a$ and $b$ in the feature space. This interpolation-based approach effectively expands the minority class distribution by creating plausible synthetic samples that preserve the original data's characteristic patterns. 

A key advantage of SMOTE is its ability to enhance minority class representation without distorting the original data distribution, thereby mitigating class imbalance issues while maintaining data authenticity \citep{chawla2002smote}. For practical implementation, we employed the SMOTE module from the imbalanced-learn Python library \footnote{https://github.com/scikit-learn-contrib/imbalanced-learn}, which provides optimized implementations of various algorithms specifically designed for handling imbalanced datasets in machine learning applications.

By combining geometric transformations with the SMOTE oversampling method, we have not only optimized the diversity and class balance of the training set, but also provided higher-quality support for the model's learning process. Specifically, the geometric transformations employed in this study (including rotation and flipping) increased the number of DLSG training samples five times, thus improving the diversity of the data set while effectively avoiding excessive repetition. Meanwhile, the use of the SMOTE algorithm to generate synthetic samples achieved a 1:1 balance between the DLSG and non-DLSG classes. This integrated approach helps mitigate potential issues caused by class imbalance while simultaneously enhancing the model's performance in both feature extraction and classification tasks.

\section{Methodology} \label{sec:method}

In classification tasks, supervised machine learning models typically rely on large amounts of high-quality labeled data to optimize the parameters of their complex models in order to improve classification accuracy and model generalization. However, for rare astronomical objects like DLSGs, which are sparsely distributed and difficult to identify in the nearby universe, the number of known samples is extremely limited. For example, in this study, the number of known DLSG samples is only 459. Therefore, although data augmentation techniques can be employed to increase the diversity of the training samples, this limited sample size is still far from meeting the requirements for labeled data volume in traditional supervised learning approaches.

To address this data bottleneck, this study employs the GC-SWGAN semi-supervised learning framework proposed by \citet{2025ApJS..279...17L}. This innovative framework integrates unsupervised generative adversarial networks with supervised learning methods, leveraging the potential information from a large number of unlabeled samples to enhance the effectiveness of feature learning and thereby significantly reduce the model's dependence on labeled samples for classification tasks. For more details on GC-SWGAN, please refer to \citet{2025ApJS..279...17L}. In this section, we will outline the framework's fundamental principles, its specific application to DLSG identification, and the model training process.

\subsection{GC-SWGAN Model for DLSGs}

GC-SWGAN is an innovative semi-supervised multi-task learning framework designed to efficiently utilize limited labeled data and abundant unlabeled data for astronomical data analysis. This framework combines semi-supervised generative adversarial networks (SGAN) with improved Wasserstein generative adversarial network (WGAN-GP), not only enhancing model convergence and stability but also significantly improving classification performance for celestial objects with scarce labels.


In this framework, the embedded discriminator and classifier are independently designed but share certain network structures and parameters. The discriminator is used to distinguish between real and generated data, while the classifier handles the classification task. Through the collaborative optimization process with the built-in generator, they not only significantly enhance the model's generative performance but also boost its classification capability.

To be specific, in GC-SWGAN, the classifier is directly trained using labeled data while sharing the feature extraction layer with the discriminator. This design allows the two components to share information within the feature space, further enhancing classification performance. Since the classifier and discriminator share part of the network architecture, optimizing the discriminator also indirectly improves the performance of the classifier. This unique collaborative optimization mechanism forms a dynamic feedback loop: improvements in the discriminator provide more accurate gradient feedback for the generator, enabling it to produce more representative synthetic samples; as the generator continues to improve through adversarial training, it enhances the discriminator’s feature extraction capability and indirectly optimizes the classification performance of the classifier. This hierarchical improvement mechanism brings significant enhancements to both data generation and classification tasks.

The network architecture of the GC-SWGAN model is illustrated in Figure 2 and primarily consists of three core components: the generator $G$, discriminator $D$, and classifier $C$. The generator $G$ comprises fully connected layers, a series of transposed convolutional layers, batch normalization layers, and Leaky ReLU activation functions. The output layer uses a tanh activation function to ensure that the pixel values of generated images fall within the range [-1, 1]. Both the discriminator $D$ and classifier $C$ are independently designed but share partial network architecture. The shared portion consists of multiple layers of CNNs, batch normalization layers, and Leaky ReLU activation functions, which are responsible for extracting multi-level features from input data. In contrast, the non-shared portion is entirely composed of fully connected layers that enable the discriminator and classifier to perform their respective specific tasks. This shared architecture not only reduces the number of model parameters, improves computational efficiency, but also allows the discriminator and classifier to synergistically optimize during the feature extraction phase, further enhancing the overall performance of the model. For detailed network structures of $D$, $C$, and $G$, please refer to the study by \citet{2025ApJS..279...17L}.

\begin{figure} 
        \includegraphics[width=\textwidth]{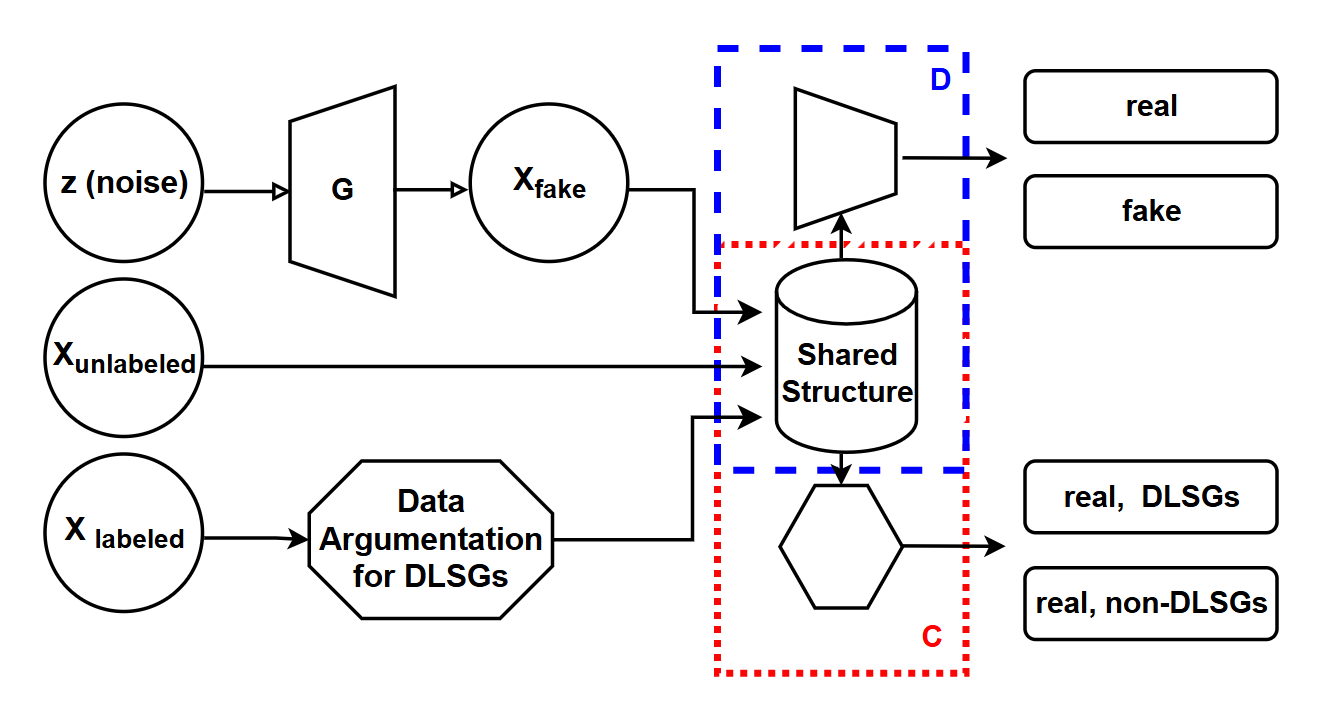} 
        \caption{Network architecture of GC-SWGAN. The symbol $ z $ represents the input random noise vector, $X_{unlabeled}$ denotes the input unlabeled real images, and $X_{labeled}$ indicates the input labeled real images, which include both DLSGs and non-DLSGs; among them, DLSG samples require data augmentation. $X_{fake}$ represents the fake images generated by the generator. $G$ signifies the generator, which employs a multi-layer convolutional network structure and is responsible for generating synthetic images that resemble real images. $D$ denotes the discriminator, whose task is to distinguish between real images and fake images. $C$ represents the classifier, which is responsible for learning to assign the correct category labels to real samples.} 
        \label{fig:framework} 
\end{figure}

In addressing the challenge of accurately identifying DLSGs in the DESI-LS, we refined the GC-SWGAN framework by optimizing both the input pipeline and the classifier’s output layer. Specifically, in terms of data input pipeline, as described in Section \ref{subsect:data_prep}, we adopted a differential augmentation strategy: for labeled DLSG samples, we increased their number through geometric transformations (including rotation and flipping) combined with SMOTE over-sampling techniques, while non-DLSG samples maintained their original data volumes. This asymmetric augmentation strategy effectively mitigated the issue of class imbalance.

For the output layer of the classifier, because the task is now binary (DLSG vs. non-DLSG), we replaced the original 10-class softmax output layer with a binary softmax layer. This modification enables the model to directly output the probability that a sample belongs to either DLSGs or non-DLSGs, with the sum of these probabilities equaling 1. This modification significantly simplifies the decision space and reduces misclassification errors caused by multi-class interference.

Experiments demonstrate that these two improvements (differential augmentation + binary classification output) enhance the model’s recognition efficiency for DLSG, significantly improving recall rates and classification accuracy compared to the original multi-class configuration, and greatly boosting overall performance. Additionally, in this study, unless otherwise specified, the classification probability threshold of the classifier is set to 0.5 by default.

\subsection{Training the Network}

During the training process, the GC-SWGAN model needs to handle three different types of input data. The first type of data is labeled real images $X_{labeled}$. In our task of identifying DLSGs, this category of data encompasses 459 DLSGs and 12,000 non-DLSG samples. The classifier $C$ will conduct supervised learning based on these annotated samples to predict whether an input sample is a DLSG or a non-DLSG.

The second type of data is unlabeled real images $X_{unlabeled}$. During the training process, only the discriminator $D$ processes this portion of the data, optimizing its feature extraction capability by judging the authenticity of the samples. Since the classifier $C$ and the discriminator $D$ share part of the network architecture, the performance of the classifier $C$ is indirectly improved during the optimization of the discriminator $D$. In this study, we use the 20,000 unlabeled DESI-LS galaxy images collected in Section \ref{subsec:sample_select} as the source of unlabeled data.

The third type of data consists of synthetic images $X_{fake}$ generated by the generator $G$. The generator $G$ produces these synthetic images through adversarial training based on an input random noise vector $z$, while the discriminator $D$ optimizes its discriminative ability by identifying these synthetic images as fake data. Although this type of data does not directly participate in the training of the classifier $C$, the optimization of the discriminator $D$ indirectly enhances the performance of the classifier $C$ due to their shared partial network architecture.

The loss function of the GC-SWGAN model consists of three components: generator loss $L_G$, discriminator loss $L_D$, and classifier loss $L_C$. Among these, the generator loss $L_G$ is an adversarial loss whose objective is to maximize the probability that the generated images are identified as real by the discriminator. We employ the Wasserstein distance to measure this probability. The discriminator loss $L_D$ consists of two components. The first part uses the Wasserstein distance to evaluate the difference between real samples and generated samples. The second part is a gradient penalty term designed to enforce Lipschitz continuity, thereby preventing issues such as gradient explosion or vanishing gradients. The classifier loss $L_C$ is a supervised loss calculated using the categorical cross-entropy function. For detailed computational procedures regarding the model's loss functions, please refer to the study by \citet{2025ApJS..279...17L}.

The model was developed using the Keras API built on TensorFlow 2. The training process was conducted on a platform equipped with an NVIDIA L40S GPU, with the batch size set to 64. The generator, discriminator, and classifier all employed the ADAM optimizer \citep{kingma2014adam} with the following parameter configuration: $\beta_1 = 0.5$ and $\beta_2 = 0.999$. The initial learning rate was set to 0.0001 and underwent exponential decay after each iteration with a decay factor of 1/1.000004. The complete model training ran for 100,000 iterations without implementing any early stopping. This fixed duration was determined empirically to ensure robust convergence and stability. The training time was approximately 20 hours to complete.

\section{Model Performance Evaluation} \label{sec:performance}

In this section,  we comprehensively evaluate the model's performance in identifying DLSGs using multiple evaluation metrics. These metrics include accuracy, recall, precision, F1-score and Matthews correlation coefficient (MCC), among others.

In general, accuracy is a widely employed metric for assessing model classification performance. It is defined as the ratio of correctly classified samples to the total number of samples, reflecting the overall correctness of the model’s predictions. The formula is:

\begin{equation}
     \text{Accuracy} = \frac{\text{TP} + \text{TN}}{\text{TP} + \text{TN} + \text{FP} + \text{FN}},
     \label{eq:accuracy} 
\end{equation}
where, TP denotes the number of true positives, TN the number of true negatives, FP the number of false positives, and FN the number of false negatives. For our trained GC-SWGAN model, the classification accuracy on the test set reached approximately 98.9\%.

However, the labeled dataset used in this study exhibits severe class imbalance: the number of DLSG samples is far smaller than that of non-DLSG samples, with DLSGs accounting for only about 3.7\% of the total. This imbalance distorts the accuracy metric - an overall high accuracy can coexist with poor classification performance on the minority class (DLSGs). In other words, the model may excel at identifying the majority class (non-DLSGs) while still performing poorly on the minority class (DLSGs), rendering accuracy an inadequate measure of the model’s true capability.

To evaluate the model’s performance on each class (DLSGs vs. non-DLSGs) more objectively, we computed additional metrics. Precision—commonly referred to as purity in the astronomical literature—measures the fraction of positively predicted samples that are actually positive, thereby reflecting the reliability of the model’s positive predictions. This metric is particularly important when false positives incur substantial costs, as it quantifies the purity of the model’s positive classifications. The formula is:

\begin{equation}
    \text{Precision} = \frac{\text{TP}}{\text{TP} + \text{FP}}.
     \label{eq:precision} 
\end{equation}

Recall—referred to as completeness in many astronomical studies—measures the proportion of actual positive samples that are correctly identified by the model, reflecting its ability to capture positive instances. This metric is especially important when evaluating performance on the minority class, as it directly quantifies the fraction of all positive cases that the model successfully detects. The calculation formula is the following:

\begin{equation}
    \text{Recall} = \frac{\text{TP}}{\text{TP} + \text{FN}}.
     \label{eq:recall} 
\end{equation}

The F1-score is the harmonic mean of precision and recall. By jointly considering these two metrics, it provides a balanced evaluation of the model's performance on positive class predictions. This measure is particularly valuable for imbalanced datasets, as it discourages overoptimizing either precision or recall alone and instead emphasizes the need for a balanced trade-off between them. Its formula is as follows:

\begin{equation}
 \text{F1-Score} = 2 \times \frac{\text{Precision} \times \text{Recall}}{\text{Precision} + \text{Recall}}.
 \label{eq:f1score} 
\end{equation}

Table \ref{tab:metrics} summarizes all classification performance metrics on the test set for both the DLSG and non-DLSG classes, along with the corresponding sample counts (support). The results demonstrate that the trained GC-SWGAN model achieves robust overall performance on the DLSG identification task. Specifically, for the majority class (non-DLSGs), the model exhibits exceptional discriminative power, with precision, recall, and F1-score all exceeding 99\%, indicating outstanding reliability. For the minority class (DLSGs), the model likewise maintains strong performance: precision reaches 84.2\%, recall 89.0\%, and F1-score 85.6\%. These outcomes confirm that, despite the pronounced class imbalance in the dataset, the model can effectively distinguish between DLSGs and non-DLSGs and attains satisfactory performance on both classes.

\begin{table}[ht]
    \centering
    \vspace{0.3cm}
    \caption{Classification performance metrics of the model for DLSGs and non-DLSGs on the test set.}
    \label{tab:metrics}
    \setlength{\tabcolsep}{10pt}
    \renewcommand{\arraystretch}{1.3}
    \begin{tabular}{rcccccc}
        \hline
        \textbf{Class} & \textbf{Precision (\%)} & \textbf{Recall (\%)} & \textbf{F1-Score (\%)} & \textbf{AU-ROC} & \textbf{AU-PRC} & \textbf{Support} \\
        \hline
        DLSGs      & 84.21 & 86.96 & 85.56 & 0.9960 & 0.8958 & 92  \\
        Non-DLSGs  & 99.50 & 99.38 & 99.44 & 0.9960 & 0.9998 & 2400 \\
        \hline
    \end{tabular}
    \vspace{0.3cm}
\end{table}

Another metric well-suited for evaluating classifiers on imbalanced datasets is the MCC. By simultaneously accounting for all four possible classification outcomes: true positives (TP), false positives (FP), true negatives (TN), and false negatives (FN), MCC provides a more comprehensive assessment of model performance. It is calculated as:

\begin{equation}
 \text{MCC} = \frac{\text{TP} \times \text{TN} - \text{FP} \times \text{FN}}{\sqrt{(\text{TP} + \text{FP})(\text{TP} + \text{FN})(\text{TN} + \text{FP})(\text{TN} + \text{FN})}}.
 \label{eq:mcc} 
\end{equation}
The range of MCC is from -1 to +1, where +1 indicates perfect prediction, -1 represents complete disagreement with the true labels, and 0 corresponds to random guessing. In our experiments, the model achieved an MCC score of 0.85 on the test set. This result demonstrates that the model maintains excellent classification performance for distinguishing between DLSGs and non-DLSGs, even when trained on severely imbalanced data.

\begin{figure} 
        \centering
        \includegraphics[width=0.8\textwidth]{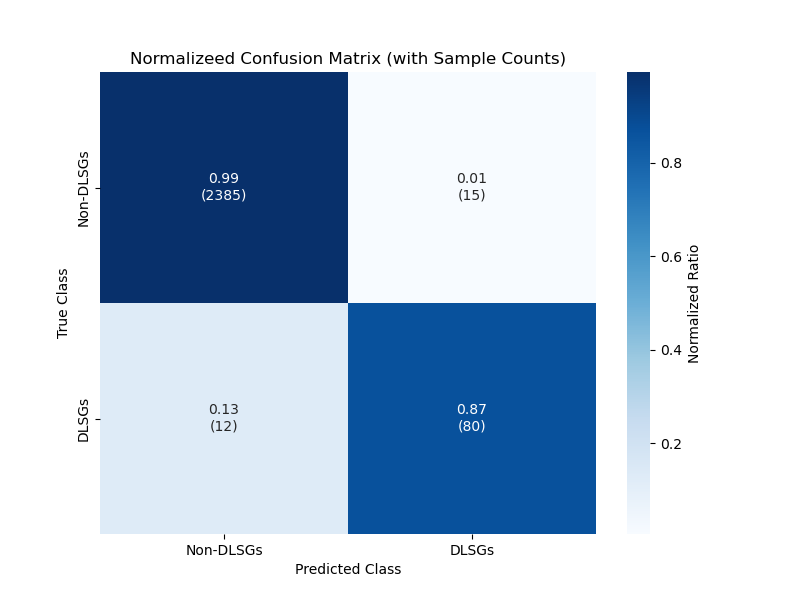} 
        \caption{Normalized confusion matrix of the model on the test set, showcasing its ability to distinguish between DLSGs and non-DLSGs. Values in the matrix denote normalized proportions; the corresponding absolute sample counts are shown in parentheses; the color bar indicates the normalized ratio.} 
        \label{fig:cm_dlsgs} 
\end{figure}

Furthermore, we analyze the model’s misclassification patterns between DLSGs and non-DLSGs using the confusion matrix shown in Figure \ref{fig:cm_dlsgs}. This matrix provides an intuitive visualization of the relationship between predicted outcomes and ground-truth labels, enabling us to pinpoint the model’s discriminative capacity for each class and to reveal any potential weaknesses.

As shown in Figure \ref{fig:cm_dlsgs}, the model excels at identifying the majority class (non-DLSGs), attaining 99\% accuracy and correctly classifying nearly every non-DLSG sample. Despite the extreme class imbalance — DLSGs constitute only 3.7\% of the data — the model demonstrates strong capability on the minority class, achieving a recall of 87\% and thus successfully detecting the majority of DLSG instances; merely 13\% of DLSGs are misclassified as non-DLSGs.

It should be emphasized that extreme class imbalance typically biases a model toward the majority class, thereby undermining its ability to recognize the minority class. In the face of this inherent challenge, our GC-SWGAN model stands out: it not only sustains 99\% accuracy on the majority class (non-DLSGs), but also attains approximately 84\% precision and 87\% recall on the minority class (DLSGs), yielding an F1-score of about 86\%. These results significantly outperform random-guess baselines and robustly demonstrate the model’s capacity to identify minority-class DLSGs under severe class imbalance.

We further introduce the area under the receiver operating characteristic curve (AU-ROC) and the area under the precision-recall curve (AU-PRC) as supplementary evaluation metrics. The AU-ROC metric quantifies model performance across all classification thresholds by measuring the area under the ROC curve, which plots the true positive rate (TPR) against the false positive rate (FPR). These rates are defined as:

\begin{equation}
 \text{TPR} = \frac{\text{TP}} {\text{TP} + \text{FN}},
 \label{eq:tpr} 
\end{equation}
\begin{equation}
 \text{FPR} = \frac{\text{FP}} {\text{TN} + \text{FP}}.
 \label{eq:fpr} 
\end{equation}
An ideal classifier achieves AU-ROC = 1, while random guessing yields AU-ROC = 0.5.

The AU-PRC metric, representing the area under the precision-recall curve, is especially valuable for imbalanced datasets (like our DLSG case) as it focuses on the model's ability to correctly identify the rare positive class. Unlike AU-ROC, which considers both positive and negative classes (through TPR and FPR), AU-PRC specifically emphasizes precision in minority class prediction. Both metrics range from 0 to 1, with higher values indicating better performance.

\begin{figure} 
        \includegraphics[width=\textwidth]{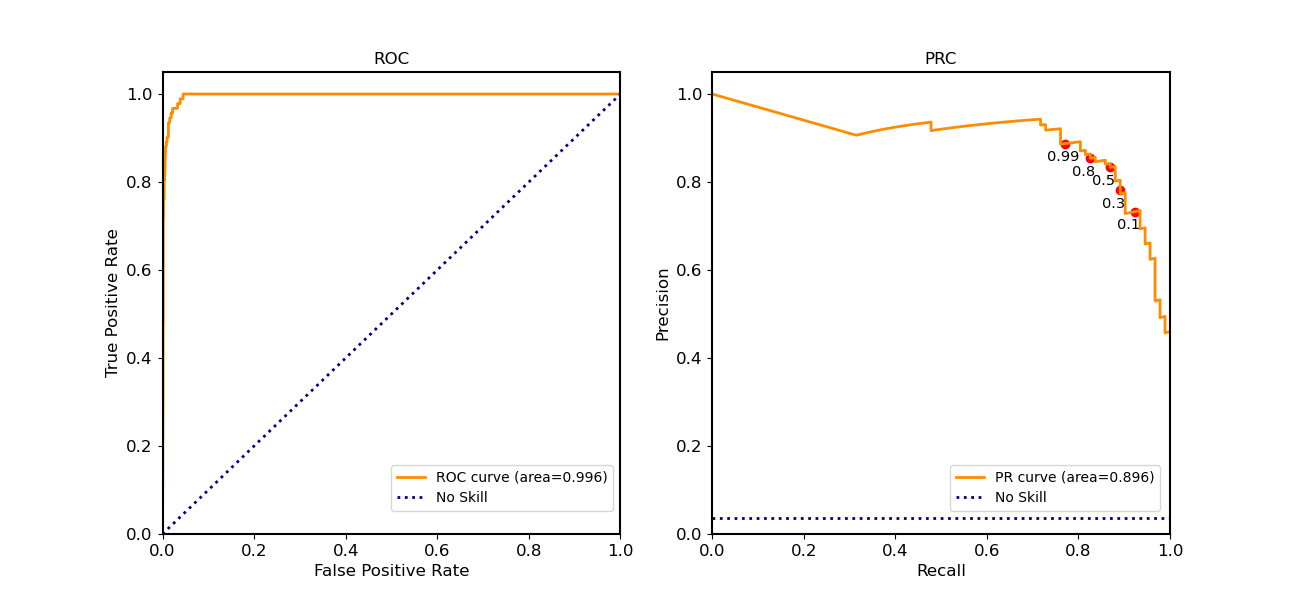} 
        \caption{Classification performance of the model for DLSGs. Left panel: The receiver operating characteristic curve (ROC) of the model on the test data. The solid orange line represents the ROC curve. The dashed diagonal line represents the performance baseline of a random classifier (AU-ROC = 0.5). Right panel: The precision-recall curve (PRC) of the model's classifier. The meanings of the lines are consistent with the left panel. The values on the curves represent different classification thresholds for DLSGs.} 
        \label{fig:ROC} 
\end{figure}

Figure \ref{fig:ROC} displays the ROC curve (left) and PRC curve (right) for the DLSG class on the test set. In the left panel, the solid orange line represents the ROC curve of the GC-SWGAN model, while the gray dashed diagonal denotes the random classifier baseline (AU-ROC = 0.5). Experimental results show that the GC-SWGAN model achieves an AU-ROC of 0.996 on the test set — substantially outperforming the random baseline — thereby confirming its outstanding discriminative capability in distinguishing DLSGs from non-DLSGs.

In the right panel, the orange solid line denotes the PRC of the GC-SWGAN model on the test set, while the horizontal dashed line marks the PRC baseline of random classification. The annotated values correspond to the classification thresholds for DLSGs. The model achieves an AU-PRC of 0.896 — substantially exceeding the random baseline. As the threshold increases, precision for DLSGs rises and recall falls, illustrating the familiar precision-recall trade-off.

Moreover, Table \ref{tab:metrics} reports  the AU-ROC and AU-PRC for both DLSGs and non-DLSGs, providing a quantitative benchmark of overall performance. These results confirm that the GC-SWGAN model maintains exceptional accuracy on the majority class (non-DLSGs) while delivering outstanding recognition capability for the minority class (DLSGs).

\section{A Catalog of DLSG Candidates in DESI-LS} \label{subsec:catalog}

The systematic evaluation in Section \ref{sec:performance} demonstrates that GC-SWGAN simultaneously maintains high accuracy on the majority class (non-DLSGs) and significantly boosts minority class (DLSGs) detection, even under severe class imbalance. These outcomes not only corroborate the model's robustness but also underscore its broad applicability to real DESI-LS imaging data. This provides both theoretical grounding and empirical support for the subsequent use of GC-SWGAN to search for DLSGs.

To construct a reliable catalog of DLSG candidates from DESI-LS, we targeted nearby bright galaxies for prediction and applied the same photometric and redshift cuts used during model training: $0.01 < z < 0.07$ and $m_r < 17.0$. The upper redshift limit, $z < 0.07$, corresponds to the maximum distance at which dust lanes can be reliably detected; \citet{2012MNRAS.423...59S} show that beyond this redshift visual identification of dust lanes becomes markedly harder and the fraction of detectable DLSGs drops sharply. The lower limit, $z > 0.01$, excludes very nearby galaxies that can suffer from incomplete image coverage, saturation, or local disturbances. The $m_r < 17.0$ magnitude cut ensures the sample’s brightness distribution matches the training data while satisfying DESI’s detection sensitivity.

\citet{walmsley2023galaxy} have released detailed morphological measurements for bright galaxies ($m_r < 19 $) within the DESI-LS footprint — covering DECaLS, MzLS, BASS, and DES — via the Zenodo platform \citep{walmsley_2023_8360385}. The catalog comprises 8.67 million galaxies and is publicly available on the Galaxy Zoo website as “Galaxy Zoo DESI (GZD).” It provides probabilities for features such as bars, spiral arms, and mergers. To facilitate use, the authors cross-matched an external, non-morphological catalog to their own; the most critical external parameter is redshift, drawn in the following order of preference: (1) SDSS spectroscopic redshifts from NSA \citep{2019ApJS..240...23A}, (2) redshifts from OSSY \citep{2015ApJS..219....1O}, and (3) spectroscopic and photometric redshifts from \citet{2021MNRAS.501.3309Z}. We adopt these redshifts in this study to ensure consistency.

From the sample of 8.67 million galaxies provided by \citet{walmsley2023galaxy}, we selected galaxies satisfying the criteria ($0.01 < z < 0.07$) and ($m_r < 17.0$), resulting in a subsample of 314,500 galaxies. Subsequently, we used the download tool \footnote{ https://www.legacysurvey.org/viewer/} provided by the DESI-LS website to obtain $g$, $r$, and $z$-band images for all these galaxies (pixel scale: 0.262 arcseconds per pixel). Based on these images, we applied the GC-SWGAN model trained in Section \ref{sec:method} for prediction to identify potential DLSGs. The model takes $g$, $r$, $z$ three-band images as input and outputs a probability $p$ of being a DLSG or $1-p$ of not being a GLSG for each galaxy. Finally, we identified the candidates for the DLSG by setting a confidence threshold.

When setting the probability threshold for distinguishing between DLSGs and non-DLSGs at 0.5, we identified a total of 9,482 DLSG candidates, representing approximately 3\% of the model’s total input sample. To our knowledge, this is currently the largest sample of DLSG candidates, with a preview of the sample catalog provided in Table 2. According to evaluation results from the test set, the precision for identifying DLSGs at this threshold is approximately 84\%, and the recall rate is around 87\%.

\begin{table}
    \centering
    \caption{Catalog of 9,482 Model-identified DLSG Candidates from the DESI Legacy Imaging Surveys. 
    \label{tab:catalog}}
    \setlength{\tabcolsep}{9.8pt} 
    \renewcommand{\arraystretch}{1.06} 
    \begin{tabular}{cccccccc}
        \hline
        \textbf{dr8\_id} & \textbf{ra}  & \textbf{dec}  & \textbf{mag\_r\_desi} & \textbf{mag\_g\_desi} & \textbf{mag\_z\_desi} & \textbf{redshift} & \textbf{prob}  \\ 
        \hline
100180\_6858 & 94.724227 & -44.139252  & 16.212643 & 17.197186 & 15.426051  & 0.067832 & 0.999063 \\
100901\_969 & 345.919455 & -44.318436  & 14.825533 & 15.859968 & 14.051870  & 0.068855 & 0.998261 \\
100958\_1289 & 5.741239 & -44.046400  & 15.786396 & 16.830297 & 14.931006  & 0.068838 & 1.000000 \\
101026\_2413 & 29.185562 & -43.972713  & 12.585146 & 13.280124 & 11.964037  & 0.015833 & 0.844298 \\
101065\_3495 & 42.651131 & -43.918285  & 15.716101 & 16.700630 & 14.894768  & 0.066280 & 0.999997 \\
101146\_1259 & 70.629257 & -44.049716  & 14.316047 & 15.188017 & 13.630199  & 0.037137 & 1.000000 \\
101847\_1333 & 313.587961 & -44.062663  & 12.056952 & 12.935183 & 11.374225  & 0.017630 & 0.979332 \\
101941\_3868 & 346.096020 & -43.922260  & 14.806158 & 15.770270 & 14.059159  & 0.049857 & 0.999326 \\
101941\_499 & 346.100138 & -44.098658  & 16.001326 & 16.920765 & 15.240787  & 0.069822 & 0.643670 \\
101993\_1927 & 3.794901 & -43.784179  & 15.047558 & 15.937565 & 14.265445  & 0.037640 & 0.862799 \\
102194\_2000 & 73.206487 & -43.781104  & 13.807968 & 14.755985 & 12.986272  & 0.021690 & 1.000000 \\
102903\_4139 & 317.738482 & -43.710882  & 13.730930 & 14.719254 & 12.942852  & 0.030000 & 0.999967 \\
102938\_41 & 329.824444 & -43.867116  & 12.699317 & 13.602638 & 11.932491  & 0.016584 & 1.000000 \\
102966\_733 & 339.384277 & -43.839358  & 14.239224 & 15.214760 & 13.513803  & 0.053918 & 0.999995 \\
102978\_1422 & 343.593716 & -43.799133  & 14.781498 & 15.782331 & 13.974561  & 0.045701 & 0.999998 \\
103093\_752 & 23.326089 & -43.542951  & 15.151112 & 16.103083 & 14.455498  & 0.059340 & 0.995484 \\
103094\_964 & 23.427981 & -43.504115  & 14.762053 & 15.734900 & 14.012280  & 0.055306 & 1.000000 \\
103168\_1205 & 48.794472 & -43.532207  & 15.730477 & 16.787586 & 14.871267  & 0.068661 & 1.000000 \\
103219\_4806 & 66.439433 & -43.405441  & 15.155093 & 16.139442 & 14.377770  & 0.051854 & 0.999964 \\
103280\_4387 & 87.470017 & -43.399081  & 13.044807 & 13.854810 & 12.401478  & 0.016849 & 0.999828 \\
103953\_5098 & 318.647634 & -43.444502  & 15.009225 & 16.025206 & 14.141737  & 0.046980 & 0.906799 \\
103957\_4589 & 319.987360 & -43.454575  & 14.640748 & 15.650880 & 13.871343  & 0.053851 & 0.996524 \\
103984\_6933 & 329.095638 & -43.407277  & 15.281905 & 16.223484 & 14.538808  & 0.050560 & 0.999997 \\
104074\_425 & 0.262028 & -43.330373  & 12.983810 & 13.906725 & 12.304114  & 0.030850 & 0.733229 \\
104074\_642 & 0.217484 & -43.333936  & 15.180485 & 16.053380 & 14.534752  & 0.040528 & 0.957892 \\
104134\_4210 & 20.762447 & -43.127860  & 13.420918 & 14.445608 & 12.518905  & 0.023820 & 1.000000 \\
104186\_2912 & 38.620073 & -43.162786  & 15.474398 & 16.442995 & 14.726337  & 0.062180 & 0.804554 \\
104245\_1180 & 58.782384 & -43.301982  & 16.336678 & 17.358097 & 15.461867  & 0.066153 & 0.765105 \\
104344\_5037 & 92.729586 & -43.167328  & 15.150101 & 16.061071 & 14.414939  & 0.042794 & 0.840935 \\
104992\_93 & 314.455981 & -43.351571  & 12.822412 & 13.689450 & 12.128301  & 0.032570 & 0.999871 \\
105019\_5576 & 323.497531 & -43.153537  & 14.679373 & 15.665434 & 13.897419  & 0.062880 & 1.000000 \\
105235\_1952 & 37.457019 & -42.962140  & 13.840979 & 14.706292 & 13.170209  & 0.019170 & 0.999999 \\
105286\_4021 & 54.689439 & -42.899930  & 14.562298 & 15.579595 & 13.733022  & 0.038773 & 1.000000 \\
106075\_2111 & 323.639554 & -43.041077  & 16.465153 & 17.651838 & 15.717813  & 0.035400 & 0.506676 \\
106226\_1094 & 15.153656 & -42.781577  & 14.995256 & 15.929481 & 14.279358  & 0.057299 & 0.999997 \\
106226\_1527 & 15.170957 & -42.750970  & 15.234915 & 16.173986 & 14.535187  & 0.053760 & 0.773159 \\
106253\_1963 & 24.391409 & -42.674005  & 13.329635 & 14.192142 & 12.647230  & 0.021900 & 0.999992 \\
106348\_472 & 56.432185 & -42.836552  & 14.997855 & 16.005737 & 14.196132  & 0.055710 & 0.715591 \\
106379\_1846 & 66.918781 & -42.777334  & 13.154992 & 13.990591 & 12.531572  & 0.022450 & 0.751991 \\
106428\_2870 & 83.858164 & -42.734540  & 16.261444 & 17.291885 & 15.392982  & 0.068490 & 0.737528 \\
106443\_1044 & 88.789260 & -42.819110  & 14.114777 & 15.046938 & 13.374405  & 0.043380 & 0.509145 \\
107116\_4739 & 317.541342 & -42.683052  & 15.504999 & 16.431427 & 14.709108  & 0.044048 & 0.997932 \\
107134\_6320 & 323.502939 & -42.648154  & 15.893241 & 17.172497 & 15.120744  & 0.028070 & 0.505756 \\

\hline
\multicolumn{6}{l}{Note: The full version of this table is available online.}
\end{tabular}
\end{table}

However, it should be noted that DLSGs account for only about 3.7\% of the test set, whereas their actual occurrence rate in observational data is likely lower, typically ranging between 1\% - 3\%. This discrepancy in class distribution may lead to slightly lower precision in DLSG detection during real-world deployment compared to the training evaluation. Nevertheless, as our later analysis shows, GC-SWGAN still demonstrates remarkable effectiveness in identifying such rare astronomical phenomena. By efficiently filtering potential candidates, the model significantly reduces the manual annotation and verification workload.

\begin{figure} 
        \includegraphics[width=\textwidth]{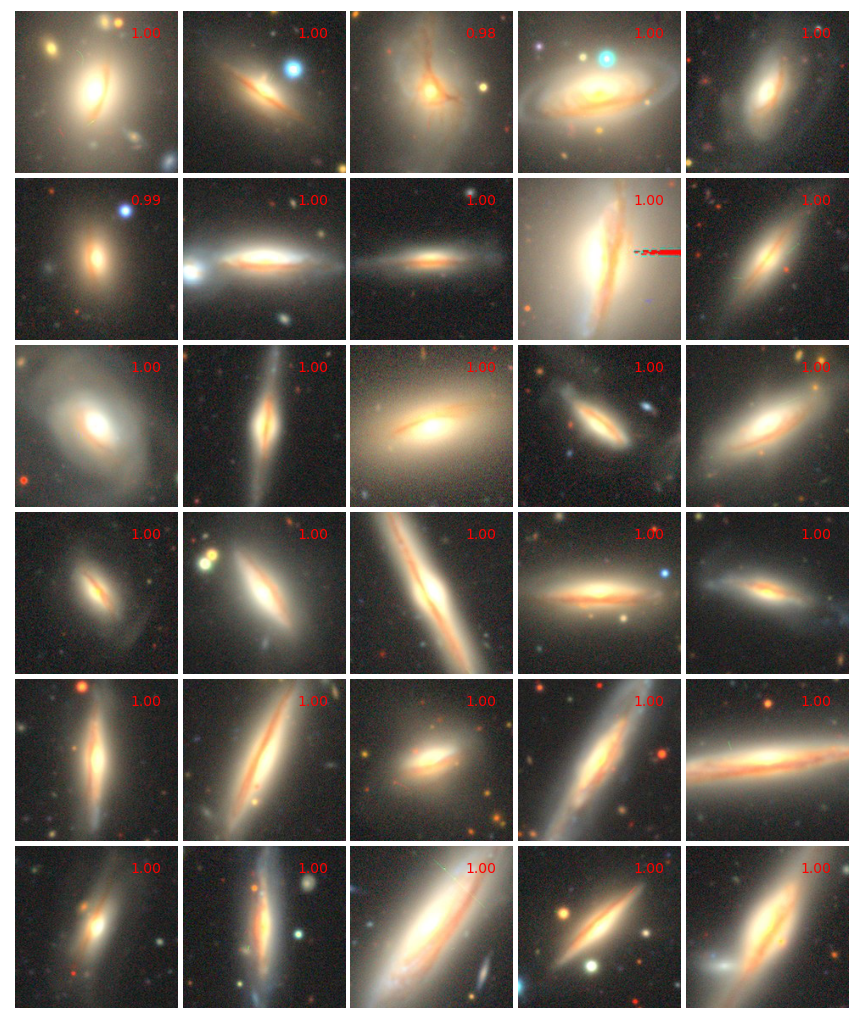} 
        \caption{ Images of 30 representative DLSGs randomly selected from the DLSG candidate catalog predicted by our model. The red numbers in each subplot indicate the probability of being predicted as a DLSG. Angular scale, data source, and color mapping follow Fig. \ref{fig:DLSGs_train}.} 
        \label{fig:dlsgs_desi} 
\end{figure}

Figure \ref{fig:dlsgs_desi} presents 30 representative early-type galaxies with prominent dust-lane features, randomly selected from the DLSG candidate catalog produced by our model. These systems all exhibit clearly visible dust-lane structures, providing direct validation of our model's reliability in identifying typical DLSGs.

\begin{figure} 
        \includegraphics[width=\textwidth]{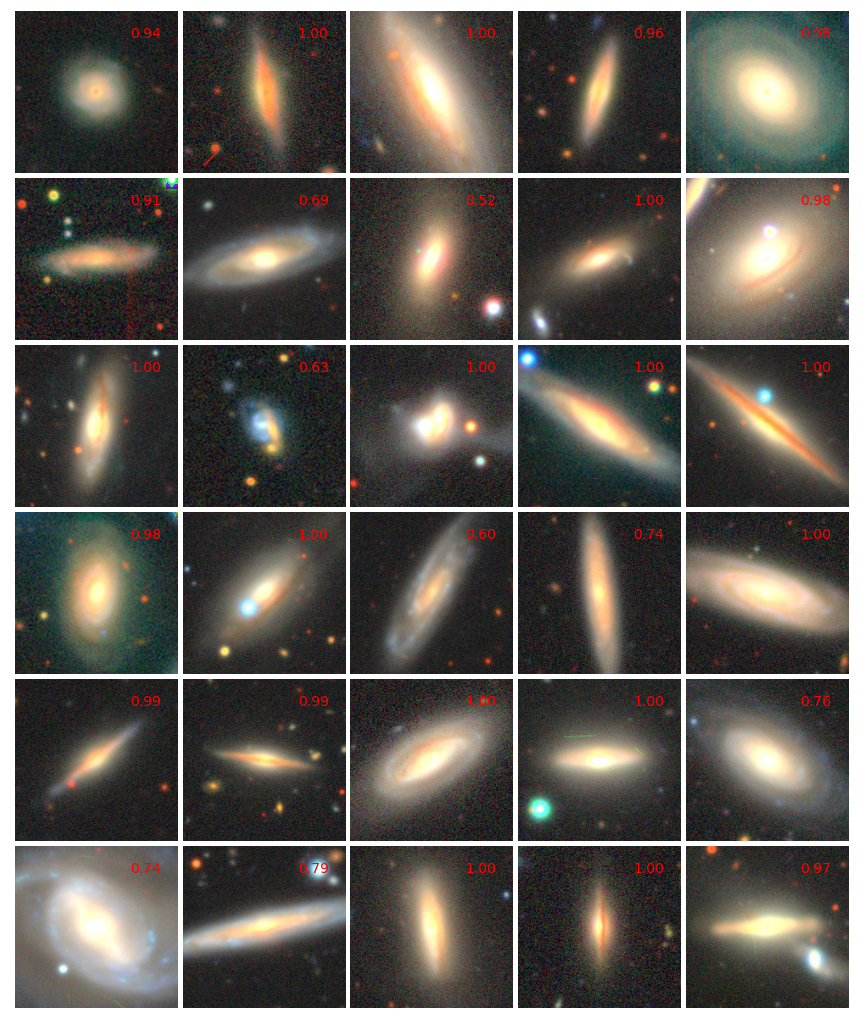} 
        \caption{ A completely random selection of 30 DLSG images from the model-predicted candidate catalog. The red number in each subplot indicates the predicted probability of being a DLSG. Angular scale, data source, and color mapping follow Fig. \ref{fig:DLSGs_train}.} 
        \label{fig:dlsgs_rand_desi} 
\end{figure}

Figure \ref{fig:dlsgs_rand_desi} shows 30 sample images that were completely randomly selected from the DLSG candidate catalog predicted by the model. These samples were selected to intuitively assess the model’s reliability. The visualization results indicate that most of the random samples exhibit clearly discernible dust lane features, suggesting that the model has a good recognition capability for typical DLSG samples. However, the analysis also reveals that a small number of samples (approximately 10\%-20\%) display fuzzy dust lane structures or unclear feature boundaries. Upon closer examination, these samples do indeed contain real but weaker dust lane features. This phenomenon indicates that the current model faces certain challenges in handling edge cases or atypical dust structure features. To enhance the model’s comprehensiveness and accuracy, future research could focus on optimizing the model’s ability to recognize these edge cases, in order to better capture and analyze various dust structure features.

Our catalogue (Table \ref{tab:catalog}) comprehensively includes all DLSG candidates identified by the model, encompassing galaxies with prominent dust lane structures as well as those with weaker features. This inclusive strategy provides researchers with a comprehensive and representative set of DLSG candidates, enabling flexible selection based on specific requirements. For instance, if high-confidence samples are needed, the classification probability threshold for DLSGs can be set to 0.9, resulting in the identification of 7,376 DLSGs (accounting for approximately 2.3\% of the original input samples). If clearer dust lane contours are desired, further examination through visual inspection by humans can be employed. Compared to directly searching in the raw data, our DLSG candidate samples provide researchers with significant time and human resource savings. By striking a balance between completeness and quality control, our catalogue meets diverse research demands and lays a solid foundation for subsequent analyses of DLSGs.

\section{Discussion} \label{sec:discuss}

Our model significantly expands the existing DLSG catalog, creating a much larger and more statistically robust sample for the systematic study of these rare galaxies. In this section, we first present the redshift and apparent magnitude distributions of the newly identified DLSGs. We then compare DLSGs with non-DLSGs using two key indicators — optical colours and star formation rates (SFRs). These comparisons reveal the distinct physical nature of DLSGs and offer preliminary clues to their evolutionary pathways. 

\subsection{Redshift and Apparent Magnitude Distributions}

\begin{figure} 
        \centering
        \includegraphics[width=0.85\textwidth]{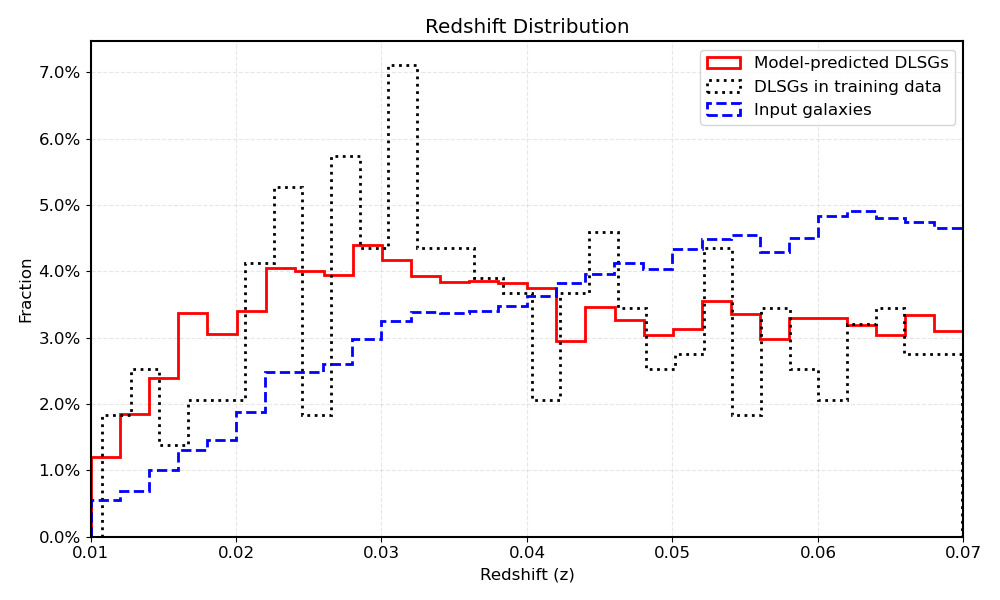} 
        \caption{Histograms of the redshift distributions for three samples: DLSGs identified by the model, DLSGs used for training, and all input galaxies. The red solid line represents the DLSGs identified by the model, the black dotted line represents the DLSGs used for training, and the blue dashed line represents all input galaxies.} 
        \label{fig:comp_redshift} 
\end{figure}

Figure \ref{fig:comp_redshift} shows the redshift distribution histograms for three samples: model-predicted DLSG samples (9,482 objects, solid red line), training set DLSG samples (459 objects, dashed black line), and total input galaxy samples (314,500 objects, dotted blue line). From the figure, it can be observed that the redshift distributions of the predicted and training set samples are highly consistent (KS test $p > 0.05$); both have a median redshift of approximately 0.04, with peaks near 0.03. However, there is a significant difference in distribution between these two samples (predicted DLSGs and training set DLSGs) and the total input galaxy sample. The KS test rejects the null hypothesis that the distributions of the predicted DLSG sample/training set DLSG sample are the same as the total sample ($p < 0.01$); the predicted DLSGs and training set DLSGs are more concentrated at lower redshifts. This result aligns with our expectations: low redshift galaxies are closer to us, making their dust lane structure features easier to observe and identify. Furthermore, since DLSGs are believed to originate from recent small-mass merger events, the enrichment of DLSGs in the low-redshift region may suggest that, during the nearby cosmic era, compact spheroidal galaxies experience minor mergers and associated gas accretion with higher probabilities.

\begin{figure} 
        \centering
        \includegraphics[width=0.85\textwidth]{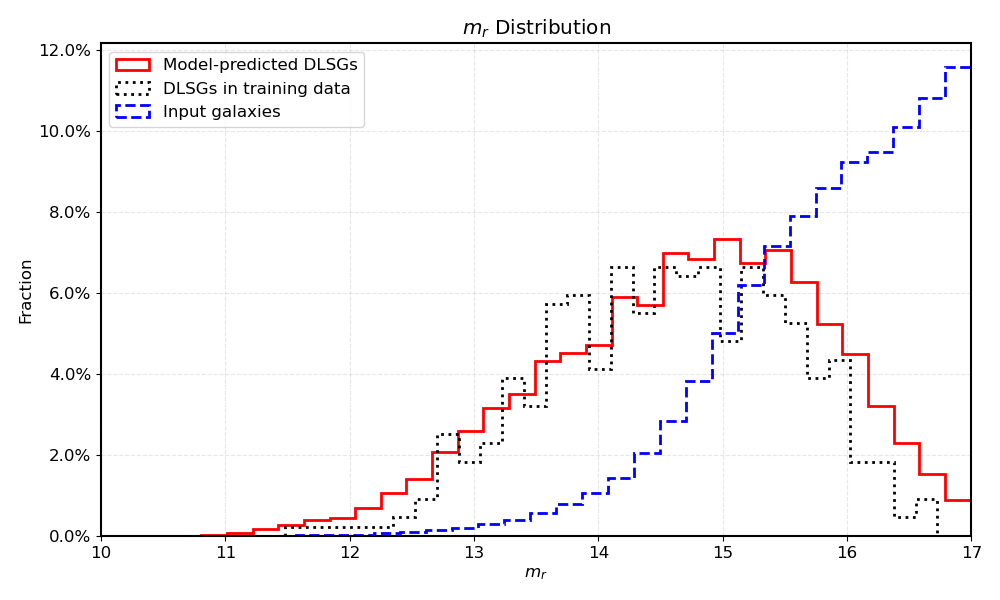} 
        \caption{Histograms of the $r$-band apparent magnitude distributions for three samples: DLSGs identified by the model, DLSGs used for training, and all input galaxies. The red solid line represents the DLSGs identified by the model, the black dotted line represents the DLSGs used for training, and the blue dashed line represents all input galaxies.} 
        \label{fig:comp_mr} 
\end{figure}

Figure \ref{fig:comp_mr} further compares the distribution of apparent magnitudes in the $r$ band for these three sample groups. Likewise, the solid red line represents the model-predicted DLSG samples, the dashed black line represents the DLSG samples from the training set, and the dotted blue line represents the total input galaxy sample. The results indicate that there is no significant difference in the $r$-band apparent magnitude distribution between the predicted samples and the training set samples (KS test $p > 0.01$), with a median value of approximately 14.5 and a peak near 14.8. However, there are significant differences between these two distributions and the total input galaxy sample (KS test $p < 0.01$): both the predicted DLSGs and the training set DLSGs are clearly concentrated in the high-luminosity region (i.e., at low apparent magnitudes). This phenomenon is also consistent with our expectations, as the morphological features of high-luminosity galaxies are typically more pronounced, making the dust lane structures easier to detect in the images.

Figure \ref{fig:comp_redshift} and \ref{fig:comp_mr} further demonstrate that the GC-SWGAN model has successfully captured the key features of DLSGs from the training dataset. The predicted DLSG samples exhibit a high degree of consistency with the training set DLSGs in terms of $r$-band apparent magnitude and redshift distribution, without being influenced by the overall input galaxy sample distribution. This indicates that the model not only accurately reproduces the core characteristics of DLSGs observed in the training data but also effectively distinguishes their unique properties from other galaxy types, thereby validating its effectiveness in identifying and predicting DLSGs.

\subsection{Color and Star Formation Properties of DLSGs}

To reliably investigate the color and star formation properties of DLSGs and quantify the impact of dust lanes on their host galaxy properties, we first performed a cross-matching between our predicted DLSG samples and two publicly available catalogs: the NASA-Sloan Atlas (NSA) catalog \citep{2019ApJS..240...23A} and the DESI photometric redshift catalog from \citet{2021MNRAS.501.3309Z}. 

We directly used the matching results from \citet{walmsley2023galaxy}. They matched the optical source coordinates from the NSA and DESI catalogs with the tractor source catalog coordinates that the DESI morphological classification catalog is based on, setting a matching radius of 10 arcseconds. In the rare cases where a single tractor source matched multiple external sources, they selected the closest external source and discarded the others. They publicly released the final matching results along with their morphological classification catalog on the GZD website \footnote{https://zenodo.org/records/4573248} for download.

Based on the cross-matching results, we successfully obtained comprehensive information for 4,214 DLSG samples, which includes critical parameters such as redshift $z$, $r$-band absolute magnitude $M_r$, concentration index $C$, $g-r$ color, and specific star formation rate $\text{Log}(\text{SFR}/M_{*})$.

Building upon this foundation, we adopted the methodologies of \citet{2007PASJ...59..541L} and \citet{2024A&A...683A..32F} to construct a control sample from approximately 300,000 non-DLSG sources. Specifically, we employed a Monte Carlo algorithm to randomly select galaxies from the non-DLSG population that match the DLSG sample in terms of redshift $z$, $r$-band absolute magnitude $M_r$, and concentration index $C$ distributions. Through these rigorous constraints, we ultimately obtained a non-DLSG control sample that is comparable in size to the DLSG sample and exhibits excellent consistency in r-band luminosity, redshift, and morphological characteristics.


To further verify the comparability between the control sample and the DLSG sample, we conducted a Kolmogorov-Smirnov (KS) test on the distributions of $z$, $M_r$, and $C$ for both samples. The test results show that all p-values are greater than 0.05, indicating that the two samples are likely drawn from the same population. Given that the colors of galaxies and the properties of star formation are closely related to stellar mass, redshift, and galaxy morphology \citep{2003MNRAS.341...33K,2003MNRAS.341...54K,2007PASJ...59..541L}, the consistency constraints of the aforementioned distributions ensure the comparability between the samples, thereby providing a reliable foundation for subsequent comparative analyses.

\begin{figure} 
        \centering
        \includegraphics[width=0.9\textwidth]{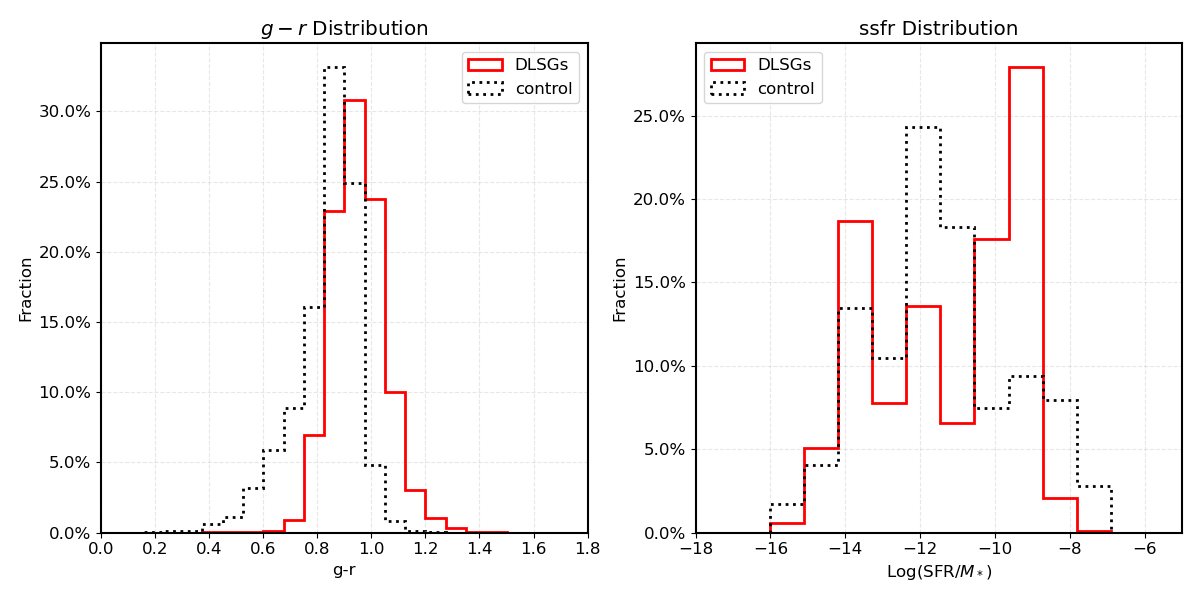} 
        \caption{Normalized distributions of $g$-$r$ color (left panel) and specific star formation rate (right panel) for DLSGs (solid red lines) and for the control sample (dotted black lines).).} 
        \label{fig:comp_color_ssfr} 
\end{figure}

Figure \ref{fig:comp_color_ssfr} illustrates the distribution differences between the DLSG sample identified by the model and the non-DLSG control sample in terms of $g-r$ color (left panel) and specific star formation rate $\text{Log}(\text{SFR}/M_*)$ (right panel). From the left panel, it is evident that DLSGs have a noticeably redder $g-r$ color (KS test $p < 0.01$), a phenomenon attributed to a more pronounced dust extinction effect in DLSGs. Specifically, due to the selective absorption of dust at different wavelengths, the radiation in the $g$ band (short wavelength) is more easily scattered and absorbed by dust particles compared to the $r$ band. This selective suppression effect leads to a preferential reduction in the $g$ band flux, resulting in a redder $g-r$ color. This statistically significant color difference not only validates the expected luminosity characteristics of dust-rich systems but also confirms our model’s capability to reliably identify galaxies exhibiting significant dust extinction effects.

The right panel further illustrates the characteristics of DLSGs in terms of star-forming activities. As can be seen from the figure, compared to the non-DLSG control samples, DLSGs exhibit a significantly higher distribution of specific star formation rates $\text{Log}(\text{SFR}/M_*)$ (KS test $p<0.01$), indicating more active star formation. Given that we have rigorously matched the DLSG and non-DLSG samples in terms of redshift, luminosity, and morphological parameters, this significant difference may reflect the unique star formation mechanisms inherent to DLSGs. Moreover, considering the prominent dust lane features commonly observed in DLSGs, we speculate that this enhanced star formation activity may originate from the reactivation of the ISM. One possible physical mechanism is that DLSGs acquire new interstellar material through minor mergers or gas accretion. These cold gases not only provide ample raw materials for star formation but also form the observed prominent dust lane structures during the dynamical processes.

Our analysis results are consistent with the findings of previous studies on DLSGs. For example, based on small sample studies, \citet{2012MNRAS.423...59S} and \citet{2012MNRAS.423...49K,2013MNRAS.435.1463K} demonstrated that DLSGs typically exhibit significant morphological disturbances and a UV-optical color biased toward blue. These features suggest active star formation activity in these galaxies and indicate signs of recent star formation. Based on these findings, Shabala and Kaviraj et al. classified these systems as starburst galaxies and further speculated that their origin may be linked to merger events involving low-mass galaxies rich in gas. Additionally, their research highlighted that the dust masses detected in these DLSGs far exceed the maximum expected values from stellar mass loss within the galaxy itself, potentially by several times. This result rules out the possibility that dust primarily originates from internal evolution of the galaxy, instead suggesting that the interstellar medium and star-forming processes may be influenced by external mechanisms. Such external influences could include gas injection, galaxy merger events, or other astrophysical phenomena.


Future research plans involve integrating higher-resolution observational data to explore the physical properties and evolutionary paths of DLSGs in greater depth, thereby providing a more comprehensive perspective for understanding the nature of these unique galaxies.

\section{SUMMARY} \label{sec:summary}

This study employs the semi-supervised deep-learning framework GC-SWGAN \citep{2025ApJS..279...17L} to conduct a systematic search for DLSGs in DESI-LS images of nearby luminous galaxies. As a distinctive type of galaxy, DLSGs have attracted considerable attention in astrophysical research due to their unique morphological features. However, traditional manual selection methods have proven inadequate for obtaining sufficient high-quality samples, thereby significantly limiting our understanding of DLSGs’ physical characteristics and evolutionary mechanisms.

To address this challenge, we carefully selected 459 rigorously reviewed DLSGs and 12,000 non-DLSGs as labeled samples, and incorporated approximately 20,000 unlabeled galaxy images for semi-supervised training. By integrating a semi-supervised generative adversarial network (SGAN) with an improved Wasserstein generative adversarial network (WGAN-GP), and applying various data augmentation techniques such as geometric transformations and SMOTE oversampling, we optimized the model’s performance in both authenticity discrimination and binary classification tasks in feature space. The results on the test set demonstrated impressive metrics: a recall rate of 86.96\%, precision of 84.21\%, an F1 score of 85.56\%, an AU-ROC of 0.996, and an AU-PRC of 0.896. These outcomes highlight the robustness of the method in tackling scenarios with extreme class imbalance.

Based on this model, we conducted predictions on 314,500 galaxy images in the DESI-LS that meet specific criteria: a redshift range of $0.01 < z < 0.07$ and an $r$-band apparent magnitude $m_r < 17.0$. By setting a probability threshold of 0.5, we identified 9,482 high-quality DLSG candidates, thereby constructing the largest and most comprehensive DLSG sample library to date. Statistical analysis of this sample reveals that the redshift and $r$-band apparent magnitude distributions of these candidates are highly consistent with those of the training samples, primarily concentrated at low redshift ($z \sim 0.03$) and high apparent brightness ($m_r \sim 15$). This suggests that nearby and high-luminosity early-type galaxies are more likely to exhibit discernible dusty lane structures in images.

We cross-matched this catalog with the NASA-Sloan Atlas \citep{2019ApJS..240...23A} and DESI photometric redshift catalogs \citep{2021MNRAS.501.3309Z}, successfully obtaining 4,214 reliable samples with complete physical parameters. By comparing these samples with a control sample constructed via Monte Carlo methods to strictly match redshift, luminosity, and morphological parameters, we found that the DLSG population exhibits significantly redder $g-r$ colors and higher specific star formation rates $\text{Log}(\text{SFR}/M_*))$ compared to non-DLSGs. These results directly corroborate the dust lanes’ absorption effects on short-wavelength light and suggest that minor mergers triggering cold gas accretion may have initiated recent star formation activity. This discovery aligns with the physical picture presented in the small-sample studies by \citet{2012MNRAS.423...59S} and \citet{2012MNRAS.423...49K,2013MNRAS.435.1463K}  while achieving significantly improved statistical significance due to the greatly expanded sample size.

In summary, this study has successfully addressed the limitations of traditional manual source selection by introducing an innovative semi-supervised GC-SWGAN framework. We have established a large-scale sample library containing about ten thousand DLSGs in the nearby universe for the first time. This achievement not only provides unprecedented statistical foundations for quantitatively studying the interactions between dust, gas, and star formation within early-type galaxies but also offers a replicable technical paradigm for survey mining other rare categories. Moving forward, our research will integrate high-resolution data resources such as JWST, ALMA, and IFU datasets to delve deeper into the physical nature of DLSGs. These advanced observations will enhance our understanding of their internal structures while providing a comprehensive perspective on their roles in cosmic evolution, thereby further advancing research in relevant fields.

\begin{acknowledgments}
Z.J.L. acknowledges the support from the National Natural Science Foundation of China (Grant No. 12573009) and the scientific research grants from the China Manned Space Project with Grand No. CMS-CSST-2025-A05 and CMS-CSST-2025-A07. S.H.Z. acknowledges support from the National Natural Science Foundation of China (Grant No. 12173026), the National Key Research and Development Program of China (Grant No. 2022YFC2807303), the Shanghai Science and Technology Fund (Grant No. 23010503900), the Program for Professor of Special Appointment (Eastern Scholar) at Shanghai Institutions of Higher Learning, and the Shuguang Program (23SG39) of the Shanghai Education Development Foundation and Shanghai Municipal Education Commission. H.B.X. acknowledges the support from the National Natural Science Foundation of China (NSFC 12203034), the Shanghai Science and Technology Fund (22YF1431500), and the Shanghai Municipal Education Commission regrading artifical intelligence empowered research. This work is also supported by the National Natural Science Foundation of China under Grant No. 12141302. 

The DESI Legacy Imaging Surveys consist of three individual projects: the Dark Energy Camera Legacy Survey (DECaLS), the Beijing-Arizona Sky Survey (BASS), and the Mayall z-band Legacy Survey (MzLS). These surveys utilized facilities such as the Blanco, Bok, and Mayall telescopes, supported by the National Science Foundation (NSF) and operated by different observatories including NSF’s NOIRLab.

We thank the respective teams and funding agencies for making these data publicly available. For detailed acknowledgments and funding information, please refer to the original publications and data release notes.
\end{acknowledgments}

\bibliography{ref}{}

\begin{thebibliography}{}
\expandafter\ifx\csname natexlab\endcsname\relax\def\natexlab#1{#1}\fi
\providecommand{\url}[1]{\href{#1}{#1}}
\providecommand{\dodoi}[1]{doi:~\href{http://doi.org/#1}{\nolinkurl{#1}}}
\providecommand{\doeprint}[1]{\href{http://ascl.net/#1}{\nolinkurl{http://ascl.net/#1}}}
\providecommand{\doarXiv}[1]{\href{https://arxiv.org/abs/#1}{\nolinkurl{https://arxiv.org/abs/#1}}}

\bibitem[{N.~K. {Agius} {et~al.}(2013){Agius}, {Sansom}, {Popescu}, {Andrae}, {Baes}, {Baldry}, {Bourne}, {Brough}, {Clark}, {Conselice}, {Cooray}, {Dariush}, {De Zotti}, {Driver}, {Dunne}, {Eales}, {Foster}, {Gomez}, {H{\"a}u{\ss}ler}, {Hopkins}, {Hopwood}, {Ivison}, {Kelvin}, {Lara-L{\'o}pez}, {Liske}, {L{\'o}pez-S{\'a}nchez}, {Loveday}, {Maddox}, {Madore}, {Phillipps}, {Robotham}, {Rowlands}, {Seibert}, {Smith}, {Temi}, {Tuffs}, \& {Valiante}}]{2013MNRAS.431.1929A}
{Agius}, N.~K., {Sansom}, A.~E., {Popescu}, C.~C., {et~al.} 2013, \bibinfo{title}{{GAMA/H-ATLAS: linking the properties of submm detected and undetected early-type galaxies - I. z {\ensuremath{\leq}} 0.06 sample},} \mnras, 431, 1929, \dodoi{10.1093/mnras/stt310}

\bibitem[{D.~S. {Aguado} {et~al.}(2019){Aguado}, {Ahumada}, {Almeida}, {Anderson}, {Andrews}, {Anguiano}, {Aquino Ort{\'\i}z}, {Arag{\'o}n-Salamanca}, {Argudo-Fern{\'a}ndez}, {Aubert}, {Avila-Reese}, {Badenes}, {Barboza Rembold}, {Barger}, {Barrera-Ballesteros}, {Bates}, {Bautista}, {Beaton}, {Beers}, {Belfiore}, {Bernardi}, {Bershady}, {Beutler}, {Bird}, {Bizyaev}, {Blanc}, {Blanton}, {Blomqvist}, {Bolton}, {Boquien}, {Borissova}, {Bovy}, {Brandt}, {Brinkmann}, {Brownstein}, {Bundy}, {Burgasser}, {Byler}, {Cano Diaz}, {Cappellari}, {Carrera}, {Cervantes Sodi}, {Chen}, {Cherinka}, {Choi}, {Chung}, {Coffey}, {Comerford}, {Comparat}, {Covey}, {da Silva Ilha}, {da Costa}, {Dai}, {Damke}, {Darling}, {Davies}, {Dawson}, {de Sainte Agathe}, {Deconto Machado}, {Del Moro}, {De Lee}, {Diamond-Stanic}, {Dom{\'\i}nguez S{\'a}nchez}, {Donor}, {Drory}, {du Mas des Bourboux}, {Duckworth}, {Dwelly}, {Ebelke}, {Emsellem}, {Escoffier}, {Fern{\'a}ndez-Trincado}, {Feuillet}, {Fischer}, {Fleming}, {Fraser-McKelvie},
  {Freischlad}, {Frinchaboy}, {Fu}, {Galbany}, {Garcia-Dias}, {Garc{\'\i}a-Hern{\'a}ndez}, {Garma Oehmichen}, {Geimba Maia}, {Gil-Mar{\'\i}n}, {Grabowski}, {Gu}, {Guo}, {Ha}, {Harrington}, {Hasselquist}, {Hayes}, {Hearty}, {Hernandez Toledo}, {Hicks}, {Hogg}, {Holley-Bockelmann}, {Holtzman}, {Hsieh}, {Hunt}, {Hwang}, {Ibarra-Medel}, {Jimenez Angel}, {Johnson}, {Jones}, {J{\"o}nsson}, {Kinemuchi}, {Kollmeier}, {Krawczyk}, {Kreckel}, {Kruk}, {Lacerna}, {Lan}, {Lane}, {Law}, {Lee}, {Li}, {Lian}, {Lin}, {Lin}, {Lintott}, {Long}, {Longa-Pe{\~n}a}, {Mackereth}, {de la Macorra}, {Majewski}, {Malanushenko}, {Manchado}, {Maraston}, {Mariappan}, {Marinelli}, {Marques-Chaves}, {Masseron}, {Masters}, {McDermid}, {Medina Pe{\~n}a}, {Meneses-Goytia}, {Merloni}, {Merrifield}, {Meszaros}, {Minniti}, {Minsley}, {Muna}, {Myers}, {Nair}, {Correa do Nascimento}, {Newman}, {Nitschelm}, {Olmstead}, {Oravetz}, {Oravetz}, {Ortega Minakata}, {Pace}, {Padilla}, {Palicio}, {Pan}, {Pan}, {Parikh}, {Parker}, {Peirani}, {Penny},
  {Percival}, {Perez-Fournon}, {Peterken}, {Pinsonneault}, {Prakash}, {Raddick}, {Raichoor}, {Riffel}, {Riffel}, {Rix}, {Robin}, {Roman-Lopes}, {Rose}, {Ross}, {Rossi}, {Rowlands}, {Rubin}, {S{\'a}nchez}, {S{\'a}nchez-Gallego}, {Sayres}, {Schaefer}, {Schiavon}, {Schimoia}, {Schlafly}, {Schlegel}, {Schneider}, {Schultheis}, {Seo}, {Shamsi}, {Shao}, {Shen}, {Shetty}, {Simonian}, {Smethurst}, {Sobeck}, {Souter}, {Spindler}, {Stark}, \& {Stassun}}]{2019ApJS..240...23A}
{Aguado}, D.~S., {Ahumada}, R., {Almeida}, A., {et~al.} 2019, \bibinfo{title}{{The Fifteenth Data Release of the Sloan Digital Sky Surveys: First Release of MaNGA-derived Quantities, Data Visualization Tools, and Stellar Library},} \apjs, 240, 23, \dodoi{10.3847/1538-4365/aaf651}

\bibitem[{F. {Annibali} {et~al.}(2010){Annibali}, {Bressan}, {Rampazzo}, {Zeilinger}, {Vega}, \& {Panuzzo}}]{2010A&A...519A..40A}
{Annibali}, F., {Bressan}, A., {Rampazzo}, R., {et~al.} 2010, \bibinfo{title}{{Nearby early-type galaxies with ionized gas. IV. Origin and powering mechanism of the ionized gas},} \aap, 519, A40, \dodoi{10.1051/0004-6361/200913774}

\bibitem[{M. Banerji {et~al.}(2010)Banerji, Lahav, Lintott, Abdalla, Schawinski, Bamford, Andreescu, Murray, Raddick, Slosar, {et~al.}}]{banerji2010galaxy}
Banerji, M., Lahav, O., Lintott, C.~J., {et~al.} 2010, \bibinfo{title}{Galaxy Zoo: reproducing galaxy morphologies via machine learning,} Monthly Notices of the Royal Astronomical Society, 406, 342, \dodoi{10.1111/j.1365-2966.2010.16713.x}

\bibitem[{E.~F. {Bell} {et~al.}(2004){Bell}, {Wolf}, {Meisenheimer}, {Rix}, {Borch}, {Dye}, {Kleinheinrich}, {Wisotzki}, \& {McIntosh}}]{2004ApJ...608..752B}
{Bell}, E.~F., {Wolf}, C., {Meisenheimer}, K., {et~al.} 2004, \bibinfo{title}{{Nearly 5000 Distant Early-Type Galaxies in COMBO-17: A Red Sequence and Its Evolution since z\raisebox{-0.5ex}\textasciitilde1},} \apj, 608, 752, \dodoi{10.1086/420778}

\bibitem[{M. {Bernardi} {et~al.}(1998){Bernardi}, {Renzini}, {da Costa}, {Wegner}, {Alonso}, {Pellegrini}, {Rit{\'e}}, \& {Willmer}}]{1998ApJ...508L.143B}
{Bernardi}, M., {Renzini}, A., {da Costa}, L.~N., {et~al.} 1998, \bibinfo{title}{{Cluster versus Field Elliptical Galaxies and Clues on Their Formation},} \apjl, 508, L143, \dodoi{10.1086/311742}

\bibitem[{M. {Bernardi} {et~al.}(2003){Bernardi}, {Sheth}, {Annis}, {Burles}, {Finkbeiner}, {Lupton}, {Schlegel}, {SubbaRao}, {Bahcall}, {Blakeslee}, {Brinkmann}, {Castander}, {Connolly}, {Csabai}, {Doi}, {Fukugita}, {Frieman}, {Heckman}, {Hennessy}, {Ivezi{\'c}}, {Knapp}, {Lamb}, {McKay}, {Munn}, {Nichol}, {Okamura}, {Schneider}, {Thakar}, \& {York}}]{2003AJ....125.1882B}
{Bernardi}, M., {Sheth}, R.~K., {Annis}, J., {et~al.} 2003, \bibinfo{title}{{Early-Type Galaxies in the Sloan Digital Sky Survey. IV. Colors and Chemical Evolution},} \aj, 125, 1882, \dodoi{10.1086/367795}

\bibitem[{R.~G. {Bower} {et~al.}(1992){Bower}, {Lucey}, \& {Ellis}}]{1992MNRAS.254..589B}
{Bower}, R.~G., {Lucey}, J.~R., \& {Ellis}, R.~S. 1992, \bibinfo{title}{{Precision photometry of early-type galaxies in the Coma and Virgo clusters : a test of the universality of the colour-magnitude relation - I. The data.},} \mnras, 254, 589, \dodoi{10.1093/mnras/254.4.589}

\bibitem[{F. {Calura} {et~al.}(2008){Calura}, {Pipino}, \& {Matteucci}}]{2008A&A...479..669C}
{Calura}, F., {Pipino}, A., \& {Matteucci}, F. 2008, \bibinfo{title}{{The cycle of interstellar dust in galaxies of different morphological types},} \aap, 479, 669, \dodoi{10.1051/0004-6361:20078090}

\bibitem[{N.~V. Chawla {et~al.}(2002)Chawla, Bowyer, Hall, \& Kegelmeyer}]{chawla2002smote}
Chawla, N.~V., Bowyer, K.~W., Hall, L.~O., \& Kegelmeyer, W.~P. 2002, \bibinfo{title}{SMOTE: synthetic minority over-sampling technique,} Journal of artificial intelligence research, 16, 321, \dodoi{10.1613/jair.953}

\bibitem[{J. Chen {et~al.}(2025)Chen, Luo, Cheng, Hou, Zhang, \& Shu}]{chen2025detecting}
Chen, J., Luo, Z., Cheng, C., {et~al.} 2025, \bibinfo{title}{Detecting Galactic Rings in the DESI Legacy Imaging Surveys with Semi-Supervised Deep Learning,} arXiv preprint arXiv:2507.07552

\bibitem[{F. {Combes} {et~al.}(2007){Combes}, {Young}, \& {Bureau}}]{2007MNRAS.377.1795C}
{Combes}, F., {Young}, L.~M., \& {Bureau}, M. 2007, \bibinfo{title}{{Molecular gas and star formation in the SAURON early-type galaxies},} \mnras, 377, 1795, \dodoi{10.1111/j.1365-2966.2007.11759.x}

\bibitem[{R.~M. {Crockett} {et~al.}(2011){Crockett}, {Kaviraj}, {Silk}, {Whitmore}, {O'Connell}, {Mutchler}, {Balick}, {Bond}, {Calzetti}, {Carollo}, {Disney}, {Dopita}, {Frogel}, {Hall}, {Holtzman}, {Kimble}, {McCarthy}, {Paresce}, {Saha}, {Trauger}, {Walker}, {Windhorst}, {Young}, {Jeong}, \& {Yi}}]{2011ApJ...727..115C}
{Crockett}, R.~M., {Kaviraj}, S., {Silk}, J.~I., {et~al.} 2011, \bibinfo{title}{{Anatomy of a Post-starburst Minor Merger: A Multi-wavelength WFC3 Study of NGC 4150},} \apj, 727, 115, \dodoi{10.1088/0004-637X/727/2/115}

\bibitem[{D.~W. {Darg} {et~al.}(2010{\natexlab{a}}){Darg}, {Kaviraj}, {Lintott}, {Schawinski}, {Sarzi}, {Bamford}, {Silk}, {Proctor}, {Andreescu}, {Murray}, {Nichol}, {Raddick}, {Slosar}, {Szalay}, {Thomas}, \& {Vandenberg}}]{2010MNRAS.401.1043D}
{Darg}, D.~W., {Kaviraj}, S., {Lintott}, C.~J., {et~al.} 2010{\natexlab{a}}, \bibinfo{title}{{Galaxy Zoo: the fraction of merging galaxies in the SDSS and their morphologies},} \mnras, 401, 1043, \dodoi{10.1111/j.1365-2966.2009.15686.x}

\bibitem[{D.~W. {Darg} {et~al.}(2010{\natexlab{b}}){Darg}, {Kaviraj}, {Lintott}, {Schawinski}, {Sarzi}, {Bamford}, {Silk}, {Andreescu}, {Murray}, {Nichol}, {Raddick}, {Slosar}, {Szalay}, {Thomas}, \& {Vandenberg}}]{2010MNRAS.401.1552D}
{Darg}, D.~W., {Kaviraj}, S., {Lintott}, C.~J., {et~al.} 2010{\natexlab{b}}, \bibinfo{title}{{Galaxy Zoo: the properties of merging galaxies in the nearby Universe - local environments, colours, masses, star formation rates and AGN activity},} \mnras, 401, 1552, \dodoi{10.1111/j.1365-2966.2009.15786.x}

\bibitem[{ {Dark Energy Survey Collaboration} {et~al.}(2016){Dark Energy Survey Collaboration}, {Abbott}, {Abdalla}, {Aleksi{\'c}}, {Allam}, {Amara}, {Bacon}, {Balbinot}, {Banerji}, {Bechtol}, {Benoit-L{\'e}vy}, {Bernstein}, {Bertin}, {Blazek}, {Bonnett}, {Bridle}, {Brooks}, {Brunner}, {Buckley-Geer}, {Burke}, {Caminha}, {Capozzi}, {Carlsen}, {Carnero-Rosell}, {Carollo}, {Carrasco-Kind}, {Carretero}, {Castander}, {Clerkin}, {Collett}, {Conselice}, {Crocce}, {Cunha}, {D'Andrea}, {da Costa}, {Davis}, {Desai}, {Diehl}, {Dietrich}, {Dodelson}, {Doel}, {Drlica-Wagner}, {Estrada}, {Etherington}, {Evrard}, {Fabbri}, {Finley}, {Flaugher}, {Foley}, {Fosalba}, {Frieman}, {Garc{\'\i}a-Bellido}, {Gaztanaga}, {Gerdes}, {Giannantonio}, {Goldstein}, {Gruen}, {Gruendl}, {Guarnieri}, {Gutierrez}, {Hartley}, {Honscheid}, {Jain}, {James}, {Jeltema}, {Jouvel}, {Kessler}, {King}, {Kirk}, {Kron}, {Kuehn}, {Kuropatkin}, {Lahav}, {Li}, {Lima}, {Lin}, {Maia}, {Makler}, {Manera}, {Maraston}, {Marshall}, {Martini}, {McMahon},
  {Melchior}, {Merson}, {Miller}, {Miquel}, {Mohr}, {Morice-Atkinson}, {Naidoo}, {Neilsen}, {Nichol}, {Nord}, {Ogando}, {Ostrovski}, {Palmese}, {Papadopoulos}, {Peiris}, {Peoples}, {Percival}, {Plazas}, {Reed}, {Refregier}, {Romer}, {Roodman}, {Ross}, {Rozo}, {Rykoff}, {Sadeh}, {Sako}, {S{\'a}nchez}, {Sanchez}, {Santiago}, {Scarpine}, {Schubnell}, {Sevilla-Noarbe}, {Sheldon}, {Smith}, {Smith}, {Soares-Santos}, {Sobreira}, {Soumagnac}, {Suchyta}, {Sullivan}, {Swanson}, {Tarle}, {Thaler}, {Thomas}, {Thomas}, {Tucker}, {Vieira}, {Vikram}, {Walker}, {Wechsler}, {Weller}, {Wester}, {Whiteway}, {Wilcox}, {Yanny}, {Zhang}, \& {Zuntz}}]{2016MNRAS.460.1270D}
{Dark Energy Survey Collaboration}, {Abbott}, T., {Abdalla}, F.~B., {et~al.} 2016, \bibinfo{title}{{The Dark Energy Survey: more than dark energy - an overview},} \mnras, 460, 1270, \dodoi{10.1093/mnras/stw641}

\bibitem[{T.~A. {Davis} \& L.~M. {Young}(2019){Davis} \& {Young}}]{2019MNRAS.489L.108D}
{Davis}, T.~A., \& {Young}, L.~M. 2019, \bibinfo{title}{{Gas accretion as fuel for residual star formation in Galaxy Zoo elliptical galaxies},} \mnras, 489, L108, \dodoi{10.1093/mnrasl/slz138}

\bibitem[{T.~A. {Davis} {et~al.}(2011){Davis}, {Alatalo}, {Sarzi}, {Bureau}, {Young}, {Blitz}, {Serra}, {Crocker}, {Krajnovi{\'c}}, {McDermid}, {Bois}, {Bournaud}, {Cappellari}, {Davies}, {Duc}, {de Zeeuw}, {Emsellem}, {Khochfar}, {Kuntschner}, {Lablanche}, {Morganti}, {Naab}, {Oosterloo}, {Scott}, \& {Weijmans}}]{2011MNRAS.417..882D}
{Davis}, T.~A., {Alatalo}, K., {Sarzi}, M., {et~al.} 2011, \bibinfo{title}{{The ATLAS$^{3D}$ project - X. On the origin of the molecular and ionized gas in early-type galaxies},} \mnras, 417, 882, \dodoi{10.1111/j.1365-2966.2011.19355.x}

\bibitem[{T.~A. {Davis} {et~al.}(2015){Davis}, {Rowlands}, {Allison}, {Shabala}, {Ting}, {Lagos}, {Kaviraj}, {Bourne}, {Dunne}, {Eales}, {Ivison}, {Maddox}, {Smith}, {Smith}, \& {Temi}}]{2015MNRAS.449.3503D}
{Davis}, T.~A., {Rowlands}, K., {Allison}, J.~R., {et~al.} 2015, \bibinfo{title}{{Molecular and atomic gas in dust lane early-type galaxies - I. Low star formation efficiencies in minor merger remnants},} \mnras, 449, 3503, \dodoi{10.1093/mnras/stv597}

\bibitem[{A. {Dey} {et~al.}(2016){Dey}, {Rabinowitz}, {Karcher}, {Bebek}, {Baltay}, {Sprayberry}, {Valdes}, {Stupak}, {Donaldson}, {Emmet}, {Hurteau}, {Abareshi}, {Marshall}, {Lang}, {Fitzpatrick}, {Daly}, {Joyce}, {Schlegel}, {Schweiker}, {Allen}, {Blum}, \& {Levi}}]{2016SPIE.9908E..2CD}
{Dey}, A., {Rabinowitz}, D., {Karcher}, A., {et~al.} 2016, in Society of Photo-Optical Instrumentation Engineers (SPIE) Conference Series, Vol. 9908, Ground-based and Airborne Instrumentation for Astronomy VI, ed. C.~J. {Evans}, L.~{Simard}, \& H.~{Takami}, 99082C, \dodoi{10.1117/12.2231488}

\bibitem[{A. {Dey} {et~al.}(2019){Dey}, {Schlegel}, {Lang}, {Blum}, {Burleigh}, {Fan}, {Findlay}, {Finkbeiner}, {Herrera}, {Juneau}, {Landriau}, {Levi}, {McGreer}, {Meisner}, {Myers}, {Moustakas}, {Nugent}, {Patej}, {Schlafly}, {Walker}, {Valdes}, {Weaver}, {Y{\`e}che}, {Zou}, {Zhou}, {Abareshi}, {Abbott}, {Abolfathi}, {Aguilera}, {Alam}, {Allen}, {Alvarez}, {Annis}, {Ansarinejad}, {Aubert}, {Beechert}, {Bell}, {BenZvi}, {Beutler}, {Bielby}, {Bolton}, {Brice{\~n}o}, {Buckley-Geer}, {Butler}, {Calamida}, {Carlberg}, {Carter}, {Casas}, {Castander}, {Choi}, {Comparat}, {Cukanovaite}, {Delubac}, {DeVries}, {Dey}, {Dhungana}, {Dickinson}, {Ding}, {Donaldson}, {Duan}, {Duckworth}, {Eftekharzadeh}, {Eisenstein}, {Etourneau}, {Fagrelius}, {Farihi}, {Fitzpatrick}, {Font-Ribera}, {Fulmer}, {G{\"a}nsicke}, {Gaztanaga}, {George}, {Gerdes}, {Gontcho}, {Gorgoni}, {Green}, {Guy}, {Harmer}, {Hernandez}, {Honscheid}, {Huang}, {James}, {Jannuzi}, {Jiang}, {Joyce}, {Karcher}, {Karkar}, {Kehoe}, {Kneib}, {Kueter-Young}, {Lan},
  {Lauer}, {Le Guillou}, {Le Van Suu}, {Lee}, {Lesser}, {Perreault Levasseur}, {Li}, {Mann}, {Marshall}, {Mart{\'\i}nez-V{\'a}zquez}, {Martini}, {du Mas des Bourboux}, {McManus}, {Meier}, {M{\'e}nard}, {Metcalfe}, {Mu{\~n}oz-Guti{\'e}rrez}, {Najita}, {Napier}, {Narayan}, {Newman}, {Nie}, {Nord}, {Norman}, {Olsen}, {Paat}, {Palanque-Delabrouille}, {Peng}, {Poppett}, {Poremba}, {Prakash}, {Rabinowitz}, {Raichoor}, {Rezaie}, {Robertson}, {Roe}, {Ross}, {Ross}, {Rudnick}, {Safonova}, {Saha}, {S{\'a}nchez}, {Savary}, {Schweiker}, {Scott}, {Seo}, {Shan}, {Silva}, {Slepian}, {Soto}, {Sprayberry}, {Staten}, {Stillman}, {Stupak}, {Summers}, {Sien Tie}, {Tirado}, {Vargas-Maga{\~n}a}, {Vivas}, {Wechsler}, {Williams}, {Yang}, {Yang}, {Yapici}, {Zaritsky}, {Zenteno}, {Zhang}, {Zhang}, {Zhou}, \& {Zhou}}]{2019AJ....157..168D}
{Dey}, A., {Schlegel}, D.~J., {Lang}, D., {et~al.} 2019, \bibinfo{title}{{Overview of the DESI Legacy Imaging Surveys},} \aj, 157, 168, \dodoi{10.3847/1538-3881/ab089d}

\bibitem[{S. Dieleman {et~al.}(2015)Dieleman, Willett, \& Dambre}]{dieleman2015rotation}
Dieleman, S., Willett, K.~W., \& Dambre, J. 2015, \bibinfo{title}{Rotation-invariant convolutional neural networks for galaxy morphology prediction,} Monthly notices of the royal astronomical society, 450, 1441, \dodoi{10.1093/mnras/stv632}

\bibitem[{K. {Ebneter} {et~al.}(1988){Ebneter}, {Djorgovski}, \& {Davis}}]{1988AJ.....95..422E}
{Ebneter}, K., {Djorgovski}, S., \& {Davis}, M. 1988, \bibinfo{title}{{A Search for Features in Early-Type Galaxies},} \aj, 95, 422, \dodoi{10.1086/114644}

\bibitem[{S.~M. {Faber} {et~al.}(1997){Faber}, {Tremaine}, {Ajhar}, {Byun}, {Dressler}, {Gebhardt}, {Grillmair}, {Kormendy}, {Lauer}, \& {Richstone}}]{1997AJ....114.1771F}
{Faber}, S.~M., {Tremaine}, S., {Ajhar}, E.~A., {et~al.} 1997, \bibinfo{title}{{The Centers of Early-Type Galaxies with HST. IV. Central Parameter Relations.},} \aj, 114, 1771, \dodoi{10.1086/118606}

\bibitem[{S.~M. {Faber} {et~al.}(2007){Faber}, {Willmer}, {Wolf}, {Koo}, {Weiner}, {Newman}, {Im}, {Coil}, {Conroy}, {Cooper}, {Davis}, {Finkbeiner}, {Gerke}, {Gebhardt}, {Groth}, {Guhathakurta}, {Harker}, {Kaiser}, {Kassin}, {Kleinheinrich}, {Konidaris}, {Kron}, {Lin}, {Luppino}, {Madgwick}, {Meisenheimer}, {Noeske}, {Phillips}, {Sarajedini}, {Schiavon}, {Simard}, {Szalay}, {Vogt}, \& {Yan}}]{2007ApJ...665..265F}
{Faber}, S.~M., {Willmer}, C.~N.~A., {Wolf}, C., {et~al.} 2007, \bibinfo{title}{{Galaxy Luminosity Functions to z\raisebox{-0.5ex}\textasciitilde1 from DEEP2 and COMBO-17: Implications for Red Galaxy Formation},} \apj, 665, 265, \dodoi{10.1086/519294}

\bibitem[{J. {Fernandez} {et~al.}(2024){Fernandez}, {Alonso}, {Mesa}, \& {Duplancic}}]{2024A&A...683A..32F}
{Fernandez}, J., {Alonso}, S., {Mesa}, V., \& {Duplancic}, F. 2024, \bibinfo{title}{{Revealing ringed galaxies in group environments},} \aap, 683, A32, \dodoi{10.1051/0004-6361/202245215}

\bibitem[{B. {Flaugher} {et~al.}(2015){Flaugher}, {Diehl}, {Honscheid}, {Abbott}, {Alvarez}, {Angstadt}, {Annis}, {Antonik}, {Ballester}, {Beaufore}, {Bernstein}, {Bernstein}, {Bigelow}, {Bonati}, {Boprie}, {Brooks}, {Buckley-Geer}, {Campa}, {Cardiel-Sas}, {Castander}, {Castilla}, {Cease}, {Cela-Ruiz}, {Chappa}, {Chi}, {Cooper}, {da Costa}, {Dede}, {Derylo}, {DePoy}, {de Vicente}, {Doel}, {Drlica-Wagner}, {Eiting}, {Elliott}, {Emes}, {Estrada}, {Fausti Neto}, {Finley}, {Flores}, {Frieman}, {Gerdes}, {Gladders}, {Gregory}, {Gutierrez}, {Hao}, {Holland}, {Holm}, {Huffman}, {Jackson}, {James}, {Jonas}, {Karcher}, {Karliner}, {Kent}, {Kessler}, {Kozlovsky}, {Kron}, {Kubik}, {Kuehn}, {Kuhlmann}, {Kuk}, {Lahav}, {Lathrop}, {Lee}, {Levi}, {Lewis}, {Li}, {Mandrichenko}, {Marshall}, {Martinez}, {Merritt}, {Miquel}, {Mu{\~n}oz}, {Neilsen}, {Nichol}, {Nord}, {Ogando}, {Olsen}, {Palaio}, {Patton}, {Peoples}, {Plazas}, {Rauch}, {Reil}, {Rheault}, {Roe}, {Rogers}, {Roodman}, {Sanchez}, {Scarpine}, {Schindler}, {Schmidt},
  {Schmitt}, {Schubnell}, {Schultz}, {Schurter}, {Scott}, {Serrano}, {Shaw}, {Smith}, {Soares-Santos}, {Stefanik}, {Stuermer}, {Suchyta}, {Sypniewski}, {Tarle}, {Thaler}, {Tighe}, {Tran}, {Tucker}, {Walker}, {Wang}, {Watson}, {Weaverdyck}, {Wester}, {Woods}, {Yanny}, \& {DES Collaboration}}]{2015AJ....150..150F}
{Flaugher}, B., {Diehl}, H.~T., {Honscheid}, K., {et~al.} 2015, \bibinfo{title}{{The Dark Energy Camera},} \aj, 150, 150, \dodoi{10.1088/0004-6256/150/5/150}

\bibitem[{M. {Fukugita} {et~al.}(2004){Fukugita}, {Nakamura}, {Turner}, {Helmboldt}, \& {Nichol}}]{2004ApJ...601L.127F}
{Fukugita}, M., {Nakamura}, O., {Turner}, E.~L., {Helmboldt}, J., \& {Nichol}, R.~C. 2004, \bibinfo{title}{{Actively Star-forming Elliptical Galaxies at Low Redshifts in the Sloan Digital Sky Survey},} \apjl, 601, L127, \dodoi{10.1086/382151}

\bibitem[{J.~S. {Gallagher} {et~al.}(2024){Gallagher}, {Kotulla}, {Laufman}, {Geist}, {Aalto}, {Falstad}, {K{\"o}nig}, {Krause}, {Privon}, {Wethers}, {Evans}, \& {Gorski}}]{2024ApJS..274....3G}
{Gallagher}, J.~S., {Kotulla}, R., {Laufman}, L., {et~al.} 2024, \bibinfo{title}{{An Imaging and Spectroscopic Exploration of the Dusty Compact Obscured Nucleus Galaxy Zw 049.057},} \apjs, 274, 3, \dodoi{10.3847/1538-4365/ad55c9}

\bibitem[{D.~H.~W. {Glass} {et~al.}(2022){Glass}, {Sansom}, {Davis}, \& {Popescu}}]{2022MNRAS.517.5524G}
{Glass}, D. H.~W., {Sansom}, A.~E., {Davis}, T.~A., \& {Popescu}, C.~C. 2022, \bibinfo{title}{{Cool interstellar medium as an evolutionary tracer in ALMA-observed local dusty early-type galaxies},} \mnras, 517, 5524, \dodoi{10.1093/mnras/stac3001}

\bibitem[{I. Goodfellow {et~al.}(2016)Goodfellow, Bengio, Courville, \& Bengio}]{goodfellow2016deep}
Goodfellow, I., Bengio, Y., Courville, A., \& Bengio, Y. 2016, Deep learning, Vol. 1, No. 2 (MIT press, Cambridge), 448

\bibitem[{P. {Goudfrooij} \& T. {de Jong}(1995){Goudfrooij} \& {de Jong}}]{1995A&A...298..784G}
{Goudfrooij}, P., \& {de Jong}, T. 1995, \bibinfo{title}{{Interstellar matter in Shapley-Ames elliptical galaxies. IV. A diffusely distributed component of dust and its effect on colour gradients.},} \aap, 298, 784, \dodoi{10.48550/arXiv.astro-ph/9504011}

\bibitem[{T.~G. {Hawarden} {et~al.}(1981){Hawarden}, {Elson}, {Longmore}, {Tritton}, \& {Corwin}}]{1981MNRAS.196..747H}
{Hawarden}, T.~G., {Elson}, R.~A.~W., {Longmore}, A.~L., {Tritton}, S.~B., \& {Corwin}, Jr., H.~G. 1981, \bibinfo{title}{{Early-type ('discledd') galaxies with dust lanes.},} \mnras, 196, 747, \dodoi{10.1093/mnras/196.4.747}

\bibitem[{K. He {et~al.}(2016)He, Zhang, Ren, \& Sun}]{he2016deep}
He, K., Zhang, X., Ren, S., \& Sun, J. 2016, in Proceedings of the IEEE conference on computer vision and pattern recognition, 770--778

\bibitem[{H. {Jeong} {et~al.}(2007){Jeong}, {Bureau}, {Yi}, {Krajnovi{\'c}}, \& {Davies}}]{2007MNRAS.376.1021J}
{Jeong}, H., {Bureau}, M., {Yi}, S.~K., {Krajnovi{\'c}}, D., \& {Davies}, R.~L. 2007, \bibinfo{title}{{Star formation and figure rotation in the early-type galaxy NGC 2974},} \mnras, 376, 1021, \dodoi{10.1111/j.1365-2966.2007.11535.x}

\bibitem[{I. {Jorgensen} {et~al.}(1996){Jorgensen}, {Franx}, \& {Kjaergaard}}]{1996MNRAS.280..167J}
{Jorgensen}, I., {Franx}, M., \& {Kjaergaard}, P. 1996, \bibinfo{title}{{The Fundamental Plane for cluster E and S0 galaxies},} \mnras, 280, 167, \dodoi{10.1093/mnras/280.1.167}

\bibitem[{G. {Kauffmann} {et~al.}(2003{\natexlab{a}}){Kauffmann}, {Heckman}, {White}, {Charlot}, {Tremonti}, {Brinchmann}, {Bruzual}, {Peng}, {Seibert}, {Bernardi}, {Blanton}, {Brinkmann}, {Castander}, {Cs{\'a}bai}, {Fukugita}, {Ivezic}, {Munn}, {Nichol}, {Padmanabhan}, {Thakar}, {Weinberg}, \& {York}}]{2003MNRAS.341...33K}
{Kauffmann}, G., {Heckman}, T.~M., {White}, S. D.~M., {et~al.} 2003{\natexlab{a}}, \bibinfo{title}{{Stellar masses and star formation histories for {}10$^{5}$ galaxies from the Sloan Digital Sky Survey},} \mnras, 341, 33, \dodoi{10.1046/j.1365-8711.2003.06291.x}

\bibitem[{G. {Kauffmann} {et~al.}(2003{\natexlab{b}}){Kauffmann}, {Heckman}, {White}, {Charlot}, {Tremonti}, {Peng}, {Seibert}, {Brinkmann}, {Nichol}, {SubbaRao}, \& {York}}]{2003MNRAS.341...54K}
{Kauffmann}, G., {Heckman}, T.~M., {White}, S. D.~M., {et~al.} 2003{\natexlab{b}}, \bibinfo{title}{{The dependence of star formation history and internal structure on stellar mass for {}10$^{5}$ low-redshift galaxies},} \mnras, 341, 54, \dodoi{10.1046/j.1365-8711.2003.06292.x}

\bibitem[{S. {Kaviraj} {et~al.}(2009){Kaviraj}, {Peirani}, {Khochfar}, {Silk}, \& {Kay}}]{2009MNRAS.394.1713K}
{Kaviraj}, S., {Peirani}, S., {Khochfar}, S., {Silk}, J., \& {Kay}, S. 2009, \bibinfo{title}{{The role of minor mergers in the recent star formation history of early-type galaxies},} \mnras, 394, 1713, \dodoi{10.1111/j.1365-2966.2009.14403.x}

\bibitem[{S. {Kaviraj} {et~al.}(2011){Kaviraj}, {Tan}, {Ellis}, \& {Silk}}]{2011MNRAS.411.2148K}
{Kaviraj}, S., {Tan}, K.-M., {Ellis}, R.~S., \& {Silk}, J. 2011, \bibinfo{title}{{A coincidence of disturbed morphology and blue UV colour: minor-merger-driven star formation in early-type galaxies at z{\ensuremath{\sim}} 0.6},} \mnras, 411, 2148, \dodoi{10.1111/j.1365-2966.2010.17754.x}

\bibitem[{S. {Kaviraj} {et~al.}(2007){Kaviraj}, {Schawinski}, {Devriendt}, {Ferreras}, {Khochfar}, {Yoon}, {Yi}, {Deharveng}, {Boselli}, {Barlow}, {Conrow}, {Forster}, {Friedman}, {Martin}, {Morrissey}, {Neff}, {Schiminovich}, {Seibert}, {Small}, {Wyder}, {Bianchi}, {Donas}, {Heckman}, {Lee}, {Madore}, {Milliard}, {Rich}, \& {Szalay}}]{2007ApJS..173..619K}
{Kaviraj}, S., {Schawinski}, K., {Devriendt}, J.~E.~G., {et~al.} 2007, \bibinfo{title}{{UV-Optical Colors As Probes of Early-Type Galaxy Evolution},} \apjs, 173, 619, \dodoi{10.1086/516633}

\bibitem[{S. {Kaviraj} {et~al.}(2008){Kaviraj}, {Khochfar}, {Schawinski}, {Yi}, {Gawiser}, {Silk}, {Virani}, {Cardamone}, {van Dokkum}, \& {Urry}}]{2008MNRAS.388...67K}
{Kaviraj}, S., {Khochfar}, S., {Schawinski}, K., {et~al.} 2008, \bibinfo{title}{{The UV colours of high-redshift early-type galaxies: evidence for recent star formation and stellar mass assembly over the last 8 billion years},} \mnras, 388, 67, \dodoi{10.1111/j.1365-2966.2008.13392.x}

\bibitem[{S. {Kaviraj} {et~al.}(2012){Kaviraj}, {Ting}, {Bureau}, {Shabala}, {Crockett}, {Silk}, {Lintott}, {Smith}, {Keel}, {Masters}, {Schawinski}, \& {Bamford}}]{2012MNRAS.423...49K}
{Kaviraj}, S., {Ting}, Y.-S., {Bureau}, M., {et~al.} 2012, \bibinfo{title}{{Galaxy Zoo: dust and molecular gas in early-type galaxies with prominent dust lanes},} \mnras, 423, 49, \dodoi{10.1111/j.1365-2966.2012.20957.x}

\bibitem[{S. {Kaviraj} {et~al.}(2013){Kaviraj}, {Rowlands}, {Alpaslan}, {Dunne}, {Ting}, {Bureau}, {Shabala}, {Lintott}, {Smith}, {Agius}, {Auld}, {Baes}, {Bourne}, {Cava}, {Clements}, {Cooray}, {Dariush}, {De Zotti}, {Driver}, {Eales}, {Hopwood}, {Hoyos}, {Ibar}, {Maddox}, {Micha{\l}owski}, {Sansom}, {Smith}, \& {Valiante}}]{2013MNRAS.435.1463K}
{Kaviraj}, S., {Rowlands}, K., {Alpaslan}, M., {et~al.} 2013, \bibinfo{title}{{A Herschel-ATLAS study of dusty spheroids: probing the minor-merger process in the local Universe},} \mnras, 435, 1463, \dodoi{10.1093/mnras/stt1629}

\bibitem[{D.~P. Kingma(2014)Kingma}]{kingma2014adam}
Kingma, D.~P. 2014, \bibinfo{title}{Adam: A method for stochastic optimization,} arXiv preprint arXiv:1412.6980

\bibitem[{G.~R. {Knapp} {et~al.}(1989){Knapp}, {Guhathakurta}, {Kim}, \& {Jura}}]{1989ApJS...70..329K}
{Knapp}, G.~R., {Guhathakurta}, P., {Kim}, D.-W., \& {Jura}, M.~A. 1989, \bibinfo{title}{{Interstellar Matter in Early-Type Galaxies. I. IRAS Flux Densities},} \apjs, 70, 329, \dodoi{10.1086/191342}

\bibitem[{G.~R. {Knapp} \& M.~P. {Rupen}(1996){Knapp} \& {Rupen}}]{1996ApJ...460..271K}
{Knapp}, G.~R., \& {Rupen}, M.~P. 1996, \bibinfo{title}{{Molecular Gas in Elliptical Galaxies: CO Observations of an IRAS Flux-limited Sample},} \apj, 460, 271, \dodoi{10.1086/176967}

\bibitem[{Z. {Luo} {et~al.}(2025a){Luo}, {Chen}, {Chen}, {Zhang}, {Fu}, {Xiao}, \& {Shu}}]{2025ApJS..279...17L}
{Luo}, Z., {Chen}, J., {Chen}, Z., {et~al.} 2025a, \bibinfo{title}{{Galaxy Morphology Classification via Deep Semisupervised Learning with Limited Labeled Data},} \apjs, 279, 17, \dodoi{10.3847/1538-4365/addb4c}

\bibitem[{Z. Luo {et~al.}(2025b)Luo, Zhang, Chen, Chen, Fu, Xiao, Du, \& Shu}]{luo2025cross}
Luo, Z., Zhang, S., Chen, J., {et~al.} 2025b, \bibinfo{title}{Cross-survey Image Transformation: Enhancing SDSS and DECaLS Images to Near-HSC Quality for Advanced Astronomical Analysis,} The Astrophysical Journal Supplement Series, 277, 22, \dodoi{10.3847/1538-4365/adaea5}

\bibitem[{Z.-J. {Luo} {et~al.}(2007){Luo}, {Shu}, \& {Huang}}]{2007PASJ...59..541L}
{Luo}, Z.-J., {Shu}, C.-G., \& {Huang}, J.-S. 2007, \bibinfo{title}{{The Differences of Star Formation History between Merging Galaxies and Field Galaxies in the Early Data Release of the SDSS},} \pasj, 59, 541, \dodoi{10.1093/pasj/59.3.541}

\bibitem[{V.~A. {Mager} {et~al.}(2018){Mager}, {Conselice}, {Seibert}, {Gusbar}, {Katona}, {Villari}, {Madore}, \& {Windhorst}}]{2018ApJ...864..123M}
{Mager}, V.~A., {Conselice}, C.~J., {Seibert}, M., {et~al.} 2018, \bibinfo{title}{{Galaxy Structure in the Ultraviolet: The Dependence of Morphological Parameters on Rest-frame Wavelength},} \apj, 864, 123, \dodoi{10.3847/1538-4357/aad59e}

\bibitem[{P. {Merluzzi}(1998){Merluzzi}}]{1998A&A...338..807M}
{Merluzzi}, P. 1998, \bibinfo{title}{{Dust in elliptical galaxies: a new dust mass evaluation},} \aap, 338, 807, \dodoi{10.48550/arXiv.astro-ph/9809147}

\bibitem[{L. {Michel-Dansac} {et~al.}(2008){Michel-Dansac}, {Lambas}, {Alonso}, \& {Tissera}}]{2008MNRAS.386L..82M}
{Michel-Dansac}, L., {Lambas}, D.~G., {Alonso}, M.~S., \& {Tissera}, P. 2008, \bibinfo{title}{{The mass-metallicity relation of interacting galaxies},} \mnras, 386, L82, \dodoi{10.1111/j.1745-3933.2008.00466.x}

\bibitem[{A.~B. {Newman} {et~al.}(2012){Newman}, {Ellis}, {Bundy}, \& {Treu}}]{2012ApJ...746..162N}
{Newman}, A.~B., {Ellis}, R.~S., {Bundy}, K., \& {Treu}, T. 2012, \bibinfo{title}{{Can Minor Merging Account for the Size Growth of Quiescent Galaxies? New Results from the CANDELS Survey},} \apj, 746, 162, \dodoi{10.1088/0004-637X/746/2/162}

\bibitem[{K. {Oh} {et~al.}(2015){Oh}, {Yi}, {Schawinski}, {Koss}, {Trakhtenbrot}, \& {Soto}}]{2015ApJS..219....1O}
{Oh}, K., {Yi}, S.~K., {Schawinski}, K., {et~al.} 2015, \bibinfo{title}{{A New Catalog of Type 1 AGNs and its Implications on the AGN Unified Model},} \apjs, 219, 1, \dodoi{10.1088/0067-0049/219/1/1}

\bibitem[{T.~A. {Oosterloo} {et~al.}(2002){Oosterloo}, {Morganti}, {Sadler}, {Vergani}, \& {Caldwell}}]{2002AJ....123..729O}
{Oosterloo}, T.~A., {Morganti}, R., {Sadler}, E.~M., {Vergani}, D., \& {Caldwell}, N. 2002, \bibinfo{title}{{Extended H I Disks in Dust Lane Elliptical Galaxies},} \aj, 123, 729, \dodoi{10.1086/338312}

\bibitem[{D. {Pandey} {et~al.}(2024){Pandey}, {Kaviraj}, {Saha}, \& {Sharma}}]{2024MNRAS.531.2223P}
{Pandey}, D., {Kaviraj}, S., {Saha}, K., \& {Sharma}, S. 2024, \bibinfo{title}{{Star formation exists in all early-type galaxies - evidence from ubiquitous structure in UV images},} \mnras, 531, 2223, \dodoi{10.1093/mnras/stae1296}

\bibitem[{S. Patro \& K.~K. Sahu(2015)Patro \& Sahu}]{patro2015normalization}
Patro, S., \& Sahu, K.~K. 2015, \bibinfo{title}{Normalization: A preprocessing stage,} arXiv preprint arXiv:1503.06462

\bibitem[{A. Radford(2015)Radford}]{radford2015unsupervised}
Radford, A. 2015, \bibinfo{title}{Unsupervised representation learning with deep convolutional generative adversarial networks,} arXiv preprint arXiv:1511.06434

\bibitem[{R.~E. {Ryan} {et~al.}(2012){Ryan}, {McCarthy}, {Cohen}, {Yan}, {Hathi}, {Koekemoer}, {Rutkowski}, {Mechtley}, {Windhorst}, {O'Connell}, {Balick}, {Bond}, {Bushouse}, {Calzetti}, {Crockett}, {Disney}, {Dopita}, {Frogel}, {Hall}, {Holtzman}, {Kaviraj}, {Kimble}, {MacKenty}, {Mutchler}, {Paresce}, {Saha}, {Silk}, {Trauger}, {Walker}, {Whitmore}, \& {Young}}]{2012ApJ...749...53R}
{Ryan}, Jr., R.~E., {McCarthy}, P.~J., {Cohen}, S.~H., {et~al.} 2012, \bibinfo{title}{{The Size Evolution of Passive Galaxies: Observations from the Wide-Field Camera 3 Early Release Science Program},} \apj, 749, 53, \dodoi{10.1088/0004-637X/749/1/53}

\bibitem[{E.~M. {Sadler} \& O.~E. {Gerhard}(1985){Sadler} \& {Gerhard}}]{1985MNRAS.214..177S}
{Sadler}, E.~M., \& {Gerhard}, O.~E. 1985, \bibinfo{title}{{How common are 'dust-lanes' in early-type galaxies ?},} \mnras, 214, 177, \dodoi{10.1093/mnras/214.2.177}

\bibitem[{L.~J. {Sage} \& G. {Galletta}(1993){Sage} \& {Galletta}}]{1993ApJ...419..544S}
{Sage}, L.~J., \& {Galletta}, G. 1993, \bibinfo{title}{{Evolution in Dust Lane Ellipticals: Detection of Accreted Molecular Gas},} \apj, 419, 544, \dodoi{10.1086/173507}

\bibitem[{S. {Salim} \& R.~M. {Rich}(2010){Salim} \& {Rich}}]{2010ApJ...714L.290S}
{Salim}, S., \& {Rich}, R.~M. 2010, \bibinfo{title}{{Star Formation Signatures in Optically Quiescent Early-type Galaxies},} \apjl, 714, L290, \dodoi{10.1088/2041-8205/714/2/L290}

\bibitem[{S.~S. {Shabala} {et~al.}(2012){Shabala}, {Ting}, {Kaviraj}, {Lintott}, {Crockett}, {Silk}, {Sarzi}, {Schawinski}, {Bamford}, \& {Edmondson}}]{2012MNRAS.423...59S}
{Shabala}, S.~S., {Ting}, Y.-S., {Kaviraj}, S., {et~al.} 2012, \bibinfo{title}{{Galaxy Zoo: dust lane early-type galaxies are tracers of recent, gas-rich minor mergers},} \mnras, 423, 59, \dodoi{10.1111/j.1365-2966.2012.20598.x}

\bibitem[{K. {Sharma} {et~al.}(2020){Sharma}, {Kembhavi}, {Kembhavi}, {Sivarani}, {Abraham}, \& {Vaghmare}}]{2020MNRAS.491.2280S}
{Sharma}, K., {Kembhavi}, A., {Kembhavi}, A., {et~al.} 2020, \bibinfo{title}{{Application of convolutional neural networks for stellar spectral classification},} \mnras, 491, 2280, \dodoi{10.1093/mnras/stz3100}

\bibitem[{F. {Shi} {et~al.}(2015){Shi}, {Liu}, {Sun}, {Li}, {Lei}, \& {Wang}}]{2015MNRAS.453..122S}
{Shi}, F., {Liu}, Y.-Y., {Sun}, G.-L., {et~al.} 2015, \bibinfo{title}{{A support vector machine for spectral classification of emission-line galaxies from the Sloan Digital Sky Survey},} \mnras, 453, 122, \dodoi{10.1093/mnras/stv1617}

\bibitem[{M.~W.~L. {Smith} {et~al.}(2012){Smith}, {Gomez}, {Eales}, {Ciesla}, {Boselli}, {Cortese}, {Bendo}, {Baes}, {Bianchi}, {Clemens}, {Clements}, {Cooray}, {Davies}, {De Looze}, {di Serego Alighieri}, {Fritz}, {Gavazzi}, {Gear}, {Madden}, {Mentuch}, {Panuzzo}, {Pohlen}, {Spinoglio}, {Verstappen}, {Vlahakis}, {Wilson}, \& {Xilouris}}]{2012ApJ...748..123S}
{Smith}, M.~W.~L., {Gomez}, H.~L., {Eales}, S.~A., {et~al.} 2012, \bibinfo{title}{{The Herschel Reference Survey: Dust in Early-type Galaxies and across the Hubble Sequence},} \apj, 748, 123, \dodoi{10.1088/0004-637X/748/2/123}

\bibitem[{M. {Sol Alonso} {et~al.}(2010){Sol Alonso}, {Michel-Dansac}, \& {Lambas}}]{2010A&A...514A..57S}
{Sol Alonso}, M., {Michel-Dansac}, L., \& {Lambas}, D.~G. 2010, \bibinfo{title}{{Metallicity of high stellar mass galaxies with signs of merger events},} \aap, 514, A57, \dodoi{10.1051/0004-6361/200912814}

\bibitem[{M. Storrie-Lombardi {et~al.}(1992)Storrie-Lombardi, Lahav, Sodre~Jr, \& Storrie-Lombardi}]{storrie1992morphological}
Storrie-Lombardi, M., Lahav, O., Sodre~Jr, L., \& Storrie-Lombardi, L. 1992, \bibinfo{title}{Morphological classification of galaxies by artificial neural networks,} Monthly Notices of the Royal Astronomical Society, 259, 8P, \dodoi{10.1093/mnras/259.1.8P}

\bibitem[{D. {Thomas} {et~al.}(1999){Thomas}, {Greggio}, \& {Bender}}]{1999MNRAS.302..537T}
{Thomas}, D., {Greggio}, L., \& {Bender}, R. 1999, \bibinfo{title}{{Constraints on galaxy formation from alpha-enhancement in luminous elliptical galaxies},} \mnras, 302, 537, \dodoi{10.1046/j.1365-8711.1999.02138.x}

\bibitem[{A. {Tomita} {et~al.}(2000){Tomita}, {Aoki}, {Watanabe}, {Takata}, \& {Ichikawa}}]{2000AJ....120..123T}
{Tomita}, A., {Aoki}, K., {Watanabe}, M., {Takata}, T., \& {Ichikawa}, S.-i. 2000, \bibinfo{title}{{The Central Gas Systems of Early-Type Galaxies Traced by Dust Features, Based on the Hubble Space Telescope WFPC2 Archival Images},} \aj, 120, 123, \dodoi{10.1086/301440}

\bibitem[{S.~C. {Trager} {et~al.}(2000{\natexlab{a}}){Trager}, {Faber}, {Worthey}, \& {Gonz{\'a}lez}}]{2000AJ....119.1645T}
{Trager}, S.~C., {Faber}, S.~M., {Worthey}, G., \& {Gonz{\'a}lez}, J.~J. 2000{\natexlab{a}}, \bibinfo{title}{{The Stellar Population Histories of Local Early-Type Galaxies. I. Population Parameters},} \aj, 119, 1645, \dodoi{10.1086/301299}

\bibitem[{S.~C. {Trager} {et~al.}(2000{\natexlab{b}}){Trager}, {Faber}, {Worthey}, \& {Gonz{\'a}lez}}]{2000AJ....120..165T}
{Trager}, S.~C., {Faber}, S.~M., {Worthey}, G., \& {Gonz{\'a}lez}, J.~J. 2000{\natexlab{b}}, \bibinfo{title}{{The Stellar Population Histories of Early-Type Galaxies. II. Controlling Parameters of the Stellar Populations},} \aj, 120, 165, \dodoi{10.1086/301442}

\bibitem[{H.~D. {Tran} {et~al.}(2001){Tran}, {Tsvetanov}, {Ford}, {Davies}, {Jaffe}, {van den Bosch}, \& {Rest}}]{2001AJ....121.2928T}
{Tran}, H.~D., {Tsvetanov}, Z., {Ford}, H.~C., {et~al.} 2001, \bibinfo{title}{{Dusty Nuclear Disks and Filaments in Early-Type Galaxies},} \aj, 121, 2928, \dodoi{10.1086/321072}

\bibitem[{A.~D. {Tubbs}(1980){Tubbs}}]{1980ApJ...241..969T}
{Tubbs}, A.~D. 1980, \bibinfo{title}{{The dynamical evolution of NGC 5128},} \apj, 241, 969, \dodoi{10.1086/158411}

\bibitem[{P.~G. {van Dokkum} \& M. {Franx}(1995){van Dokkum} \& {Franx}}]{1995AJ....110.2027V}
{van Dokkum}, P.~G., \& {Franx}, M. 1995, \bibinfo{title}{{Dust in the Cores of Early-Type Galaxies},} \aj, 110, 2027, \dodoi{10.1086/117667}

\bibitem[{M. Walmsley {et~al.}(2020)Walmsley, Smith, Lintott, Gal, Bamford, Dickinson, Fortson, Kruk, Masters, Scarlata, {et~al.}}]{walmsley2020galaxy}
Walmsley, M., Smith, L., Lintott, C., {et~al.} 2020, \bibinfo{title}{Galaxy Zoo: probabilistic morphology through Bayesian CNNs and active learning,} Monthly Notices of the Royal Astronomical Society, 491, 1554, \dodoi{10.1093/mnras/stz2816}

\bibitem[{M. {Walmsley} {et~al.}(2022){Walmsley}, {Scaife}, {Lintott}, {Lochner}, {Etsebeth}, {G{\'e}ron}, {Dickinson}, {Fortson}, {Kruk}, {Masters}, {Mantha}, \& {Simmons}}]{2022MNRAS.513.1581W}
{Walmsley}, M., {Scaife}, A. M.~M., {Lintott}, C., {et~al.} 2022, \bibinfo{title}{{Practical galaxy morphology tools from deep supervised representation learning},} \mnras, 513, 1581, \dodoi{10.1093/mnras/stac525}

\bibitem[{M. Walmsley {et~al.}(2023{\natexlab{a}})Walmsley, G{\'e}ron, Kruk, Scaife, Lintott, Masters, Dawson, Dickinson, Fortson, Garland, {et~al.}}]{walmsley2023galaxy}
Walmsley, M., G{\'e}ron, T., Kruk, S., {et~al.} 2023{\natexlab{a}}, \bibinfo{title}{Galaxy Zoo DESI: Detailed morphology measurements for 8.7 M galaxies in the DESI Legacy Imaging Surveys,} Monthly Notices of the Royal Astronomical Society, 526, 4768, \dodoi{10.1093/mnras/stad2919}

\bibitem[{M. Walmsley {et~al.}(2023{\natexlab{b}})Walmsley, Géron, Kruk, Scaife, Lintott, Masters, Dawson, Dickinson, Fortson, Garland, Mantha, O'Ryan, Popp, Simmons, Baeten, \& Macmillan}]{walmsley_2023_8360385}
Walmsley, M., Géron, T., Kruk, S., {et~al.} 2023{\natexlab{b}}, \bibinfo{title}{Galaxy Zoo DESI: Detailed Morphology Classifications for 8.7M Galaxies in the DESI Legacy Imaging Surveys,}, 1.0.1 Zenodo, \dodoi{10.5281/zenodo.8360385}

\bibitem[{L.-L. {Wang} {et~al.}(2023){Wang}, {Zheng}, {Rong}, {Yang}, {Zhang}, {Xie}, {Wang}, \& {Zhao}}]{2023NewA...9901965W}
{Wang}, L.-L., {Zheng}, W.-Y., {Rong}, L.-X., {et~al.} 2023, \bibinfo{title}{{Spectral classification of LAMOST emission line galaxies based on machine learning methods},} \na, 99, 101965, \dodoi{10.1016/j.newast.2022.101965}

\bibitem[{K.~W. Willett {et~al.}(2013)Willett, Lintott, Bamford, Masters, Simmons, Casteels, Edmondson, Fortson, Kaviraj, Keel, {et~al.}}]{willett2013galaxy}
Willett, K.~W., Lintott, C.~J., Bamford, S.~P., {et~al.} 2013, \bibinfo{title}{Galaxy Zoo 2: detailed morphological classifications for 304 122 galaxies from the Sloan Digital Sky Survey,} Monthly Notices of the Royal Astronomical Society, 435, 2835, \dodoi{10.1093/mnras/stt1458}

\bibitem[{G.~G. {Williams} {et~al.}(2004){Williams}, {Olszewski}, {Lesser}, \& {Burge}}]{2004SPIE.5492..787W}
{Williams}, G.~G., {Olszewski}, E., {Lesser}, M.~P., \& {Burge}, J.~H. 2004, in Society of Photo-Optical Instrumentation Engineers (SPIE) Conference Series, Vol. 5492, Ground-based Instrumentation for Astronomy, ed. A.~F.~M. {Moorwood} \& M.~{Iye}, 787--798, \dodoi{10.1117/12.552189}

\bibitem[{S.~K. {Yi} {et~al.}(2005){Yi}, {Yoon}, {Kaviraj}, {Deharveng}, {Rich}, {Salim}, {Boselli}, {Lee}, {Ree}, {Sohn}, {Rey}, {Lee}, {Rhee}, {Bianchi}, {Byun}, {Donas}, {Friedman}, {Heckman}, {Jelinsky}, {Madore}, {Malina}, {Martin}, {Milliard}, {Morrissey}, {Neff}, {Schiminovich}, {Siegmund}, {Small}, {Szalay}, {Jee}, {Kim}, {Barlow}, {Forster}, {Welsh}, \& {Wyder}}]{2005ApJ...619L.111Y}
{Yi}, S.~K., {Yoon}, S.~J., {Kaviraj}, S., {et~al.} 2005, \bibinfo{title}{{Galaxy Evolution Explorer Ultraviolet Color-Magnitude Relations and Evidence of Recent Star Formation in Early-Type Galaxies},} \apjl, 619, L111, \dodoi{10.1086/422811}

\bibitem[{L.~M. {Young} {et~al.}(2011){Young}, {Bureau}, {Davis}, {Combes}, {McDermid}, {Alatalo}, {Blitz}, {Bois}, {Bournaud}, {Cappellari}, {Davies}, {de Zeeuw}, {Emsellem}, {Khochfar}, {Krajnovi{\'c}}, {Kuntschner}, {Lablanche}, {Morganti}, {Naab}, {Oosterloo}, {Sarzi}, {Scott}, {Serra}, \& {Weijmans}}]{2011MNRAS.414..940Y}
{Young}, L.~M., {Bureau}, M., {Davis}, T.~A., {et~al.} 2011, \bibinfo{title}{{The ATLAS$^{3D}$ project - IV. The molecular gas content of early-type galaxies},} \mnras, 414, 940, \dodoi{10.1111/j.1365-2966.2011.18561.x}

\bibitem[{R. {Zhou} {et~al.}(2021){Zhou}, {Newman}, {Mao}, {Meisner}, {Moustakas}, {Myers}, {Prakash}, {Zentner}, {Brooks}, {Duan}, {Landriau}, {Levi}, {Prada}, \& {Tarle}}]{2021MNRAS.501.3309Z}
{Zhou}, R., {Newman}, J.~A., {Mao}, Y.-Y., {et~al.} 2021, \bibinfo{title}{{The clustering of DESI-like luminous red galaxies using photometric redshifts},} \mnras, 501, 3309, \dodoi{10.1093/mnras/staa3764}

\end{thebibliography}
\bibliographystyle{aasjournalv7}



\end{document}